\documentstyle[12pt]{article}
\begin{document}
\renewcommand{\theequation}{\thesection.\arabic{equation}}

\def\a{\alpha}
\def\b{\beta}
\def\ch{\chi}
\def\d{\delta}
\def\e{\epsilon}
\def\E{{\cal E}}
\def\f{\phi}
\def\g{\gamma}
\def\h{\eta}
\def\i{\iota}
\def\j{\psi}
\def\k{\kappa}
\def\l{\lambda}
\def\m{\mu}
\def\n{\nu}
\def\o{\omega}
\def\p{\pi}
\def\q{\theta}
\def\r{\rho}
\def\s{\sigma}
\def\t{\tau}
\def\u{\upsilon}
\def\x{\xi}
\def\z{\zeta}
\def\D{\Delta}
\def\F{\Phi}
\def\G{\Gamma}
\def\J{\Psi}
\def\L{\Lambda}
\def\O{\Omega}
\def\P{\Pi}
\def\S{\Sigma}
\def\U{\Upsilon}
\def\X{\Xi} 
\def\T{\Theta}
\def\vf{\varphi}
\def\ve{\varepsilon}
\def\cC{{\cal P}}
\def\cD{{\cal Q}}

\def\Ab{\bar{A}}
\def\gi{g^{-1}}
\def\li{{ 1 \over \l } }
\def\lb{\l^{*}}
\def\zb{\bar{z}}
\def\ub{u^{*}}
\def\vb{v^{*}}
\def\Tb{\bar{T}}
\def\pp {\partial }
\def\pb {\bar{\partial }}
\def\be{\begin{equation}}
\def\ee{\end{equation}}
\def\ben{\begin{eqnarray}}
\def\een{\end{eqnarray}}
\def\lt{\tilde{\lambda}}
\addtolength{\topmargin}{-0.3in}
\addtolength{\textheight}{1in}
\vsize=26truecm

\thispagestyle{empty}
\begin{flushright} May \ 1996\\
SNUTP 96-034 \\
hep-th/9605052 \\
\end{flushright}
\begin{center}
 {\large\bf Effective Field Theory for \\
  Coherent Optical Pulse Propagation  }
\vglue .5in
 Q-Han Park\footnote{ E-mail address; qpark@nms.kyunghee.ac.kr }
\vglue .2in
{and}
\vglue .2in
H. J. Shin\footnote{ E-mail address; hjshin@nms.kyunghee.ac.kr }
\vglue .2in
{\it  
Department of Physics \\
and \\
Research Institute of Basic Sciences \\
Kyunghee University\\
Seoul, 130-701, Korea}
\vglue .2in
{\bf ABSTRACT}\\[.2in]
\end{center}
Hidden nonabelian symmetries in nonlinear interactions of radiation 
with matter are clarified. In terms of a nonabelian potential variable, 
we construct an effective field theory of self-induced transparency, 
a phenomenon of lossless coherent pulse propagation, in association 
with Hermitian symmetric spaces $G/H$. Various new properties of 
self-induced transparency, e.g. soliton numbers, effective potential 
energy, gauge symmetry and discrete symmetries, modified pulse area, 
conserved $U(1)$-charge etc. are addressed and elaborated in the 
nondegenerate two-level case where $G/H = SU(2)/U(1)$. Using the 
$U(1)$-charge conservation, a new type of analysis on pulse stability 
is given which agrees with earlier numerical results.
\vglue .1in
\newpage
\tableofcontents
\section{Introduction}
\setcounter{equation}{0}
Since the invention of the laser, much progress has been made in 
understanding nonlinear interactions of radiation with matter which 
made nonlinear optics a fast developing and independent field of 
science. Laser light in general is expressed in terms of a 
macroscopic, classical electric field which interacts with 
microscopic, quantum mechanical matter. Unlike the case of classical 
electrodynamics, the electric scalar potential and the magnetic 
vector potential do not appear to replace electromagnetic fields 
in nonlinear optics. Instead, the electric field itself, with 
appropriate restrictions to accomodate specific physical systems, 
plays the role of a fundamental variable which renders the problem 
lacking a field theoretic Lagrangian formulation. However, one notable 
exception is the area theorem introduced by McCall and 
Hahn \cite{McCall}. In 1967, McCall and Hahn have discovered a 
remarkable coherent mode of lossless light pulses which propagate in 
a resonant, nondegenerate two-level atomic medium with inhomogeneous 
broadening, and named the phenomenon as self-induced 
transparency (SIT). In their work, propagation of light pulses is 
depicted in terms of a potential-like variable $\theta  (x)$, the 
time area of a suitably chosen electric field, which is determined 
according to the area theorem. Under certain circumstances, the system 
can be described by an effective potential variable $\vf (x,t)$ which 
satisfies the well-known sine-Gordon field theory equation. In this 
case, the 1-soliton of the sine-Gordon theory is identified with the 
$2\pi $ pulse of McCall and Hahn. The cosine potential term becomes 
proportional to the microscopic atomic energy and accounts for pulse 
stability. However, one serious drawback of the sine-Gordon 
approach is its oversimplification. In the sine-Gordon limit,  
frequency detuning and frequency modulation effects are all ignored 
and microscopic atomic motions (inhomogeneous broadening) are
not taken into account. Also, the model is limited to the nondegenerate 
two-level case whereas many interesting physical systems are degenerate 
and/or multi-level systems. Until now, only the sine-Gordon theory was 
a known field theory for SIT. Thus, subsequent generalizations of SIT 
to more complex systems have resorted only to the SIT equation, i.e. the 
Maxwell-Bloch equation under slowly varying envelope approximation 
(SVEA), finding soliton type solutions by the inverse scattering method 
in some cases. It should be noted that even though the SIT equation 
expressed in terms of a $U-V$ pair in the inverse scattering formalism 
resembles a nonlinear sigma model equation, it does not provide a field 
theory like the sine-Gordon theory whose equation of motion becomes the 
SIT equation \cite{maim2}. Following the pioneering work of 
Lamb \cite{lamb}, Ablowitz, Kaup and Newell have extended the inverse 
scattering formalism to include inhomogeneous broadening and obtained 
exact solutions \cite{AKN}. In accordance with the area theorem, these 
solutions show that an arbitrary initial pulse with sufficient strength 
decomposes into a finite number of $2\pi $ pulses and $0\pi $  pulses, 
plus radiation which decays exponentially. Extensions to the degenerate 
as well as the multi-level cases have been also found with more 
complicated soliton solutions \cite{maim2,maim,bash1,Bash2}. 

The purpose of this paper is to clarify nonabelian group structures 
and their physical properties hidden in nonlinear interactions of 
radiation with matter. These group structures allow us to construct 
an effective field theory of SIT beyond the sine-Gordon limit. 
In fact, without inhomogeneous broadening all generalizations of 
the SIT equation which admit the inverse scattering formalism can 
be described by group theoretic effective field theories which 
generalize the sine-Gordon theory to the nonabelian cases. We extend 
the potential variable $\vf $ to a matrix valued scalar potential $g$ 
and construct a field theory of SIT in terms of a certain two 
dimensional nonlinear sigma model associated with a particular coset 
$G/H$ for a pair of Lie groups $G$ and $H$. In order to do so, we 
first find a group theoretic formulation of the Bloch equation in 
terms of the motion of a spinning top moving along the coadjoint 
orbit which is determined by $G/H$, and then express the full 
Maxwell-Bloch equation under SVEA as a field theory generalization 
of the spinning top. We make such a field theory generalization by 
modifying the $G/H$ gauged Wess-Zumino-Witten nonlinear sigma model 
action which is a well-known model in mathematical physics. The 
$SU(2)/U(1)$ case, being identified with the so-called complex 
sine-Gordon theory, provides a field theory for the nondegenerate 
two-level SIT. Frequency detuning and frequency modulation effects 
are taken into account through the $SU(2)$ valued scalar potential 
$g$, or stated differently, through three scalar potential functions. 
The simplest, sine-Gordon limit arises from the restriction of $SU(2)$ 
to $U(1) \times U(1)$ and our spinning top interpretation reduces to 
the well-known physical interpretation of the sine-Gordon equation as 
the equation of motion for a chain of pendulums. Other generalizations 
of SIT to the multi-level and the degenerate cases can be made in terms 
of more complicated coset structures. It is shown that all these 
generalizations correspond to specific cosets $G/H$ known as Hermitian 
symmetric spaces. In the presense of inhomogeneous broadening, we still 
have the concept of effective potential where the matrix scalar 
potential $g$ is also a function of detuning frequency. That is, we 
have infinitely many microscopic potential functions $g$ labeled by 
detuning frequency. This prevents SIT with inhomogeneous broadening 
from a field theory formulation. Certain field theoretic properties, 
e.g. conservation laws, no longer hold in this case which nevertheless 
receive interesting physical interpretations as discussed in Sec. 6. 
The microscopic potential $g$ also reveals the group structure of the 
dressing method, a solution finding technique equvialent to the inverse 
scattering.

The effective field theory description of SIT provides a completely 
new angle to the SIT problem and reveals various noble aspects of SIT 
which were not available in the inverse scattering approach. One marked 
difference is the appearance of a group theoretical potential energy 
term which clarifies the topological and the nontopological nature of 
optical pulses. In particular, the periodicity of the potential results 
in infinitely many degenerate vacua. In order for optical pulses to 
possess finite energy, they have to reach one of the degenerate vacua 
asymptotically as $x \rightarrow \pm \infty $ which assigns specific 
soliton numbers, more than one in some cases, to each pulses.
We present an explicit form of the potential energy term for various 
multi-level and degenerate cases of SIT and define soliton numbers 
for each cases. The full Lagrangian itself also reveals new types of 
symmetries of SIT. It possesses two kinds of symmetries, first the 
continuous type; the local $H$ vector gauge symmetry and the global 
$U(1)$ axial vector gauge symmetry, and secondly the discrete type; 
the Krammers-Wannier type duality and the chiral symmetry. It is 
noted that the SIT equation is not invariant under the vector gauge 
transformation and the specific choice of gauge fixing incorporates 
frequency detuning effect. This shows that inhomogeneous broadening  
is equivalent to ``averaging" over different gauge fixings. Faraday 
rotation in the presense of external magnetic field is also 
identified with the vector gauge transformation. Discrete symmetries 
generate new solutions from a known one. In particular, the 
Krammers-Wannier type duality relates the ``bright" soliton 
with the ``dark" soliton of SIT. 
Our group theoretic formulation of SIT allows a systematic 
understanding of the integrability of SIT. We find infinitely many 
conserved local integrals of SIT using the properties of Hermitian 
symmetric spaces. The group structure of SIT also allows us to find 
exact soliton solutions using the dressing method for SIT.

The topological nature of solitons is particularly useful in 
understanding the stability of optical pulses. Due to the infinite 
energy barrier, topological solitons are stable against topological 
number changing fluctuations. In the case of nontopological solitons, 
we show that the $U(1)$ axial charge replaces the topological number. 
By making a perturbative analysis around topological and nontopological 
solitons and also using the $U(1)$ charge conservation, we find the 
stability behavior of solitons against small fluctuations. This result 
agrees well with earlier numerical work and provides a systematic 
stability analysis based on a field theory formulation which otherwise 
would not have been possible.

The plan of the paper is the following; in Sec. 2, we introduce a 
semi-classical Maxwell-Bloch description of the optical pulse 
propagation problem and the McCall and Hahn's area theorem. The 
effective potential concept is briefly discussed. In Sec. 3, we 
construct an effective field theory of SIT using a field theory 
generalization of the spinning top equation. Various cases of SIT 
are associated with specific cosets. Inhomogenous broadening is 
introduced from our field theory point of view and we explain the 
notion of effective potential energy and its degenerate vacua and 
associated soliton numbers. In Sec. 4, the dressing method is explained 
and applied to obtain soliton solutions for the nondegenerate two-level 
and the degenerate three-level system. In Sec. 5, infinitely many 
conserved local integrals are obtained using the properties of 
Hermitian symmetric spaces. Discrete and continuous symmetries of 
effective theories are aslo explained. The issue of stability is 
addressed in Sec. 6, particularly in regard to the conserved $U(1)$ 
charge and inhomogeneous broadening effect is addressed. 
Finally, Sec. 7 is a discussion.

\section{Self-induced transparency}
\setcounter{equation}{0}
The multi-mode optical pulses propagating in a resonant medium along 
the $x$-axis are described by the electric field of the form,
\be
{ \bf E} = \sum_{l=1}^{m}{\cal {\bf E}}_{l}(x, t) \exp i( k_{l}x - 
w_{l}t ) + \mbox{c.c.}
\label{elec} 
\ee
where $k_{l}$ and $w_{l}$ denote the wave number and the frequency 
of each mode and the amplitude vector $ {\cal {\bf E}}_{l} $ is in 
general a complex vector function. The governing equation of propagation 
is the Maxwell equation,
\be
({\pp^{2} \over \pp x^{2}} - {n^{2}\over c^{2}}{\pp^{2} \over 
\pp t^{2}} )
{\bf E} = {4\pi \over c^{2}}{\pp^{2} \over \pp t^{2}}\int dv 
\mbox{ tr } \rho {\bf d} .
\label{maxwell}
\ee
On the right hand side, electric dipole transitions are treated 
semiclassically. ${\bf d}$ is the atom's dipole moment operator and 
the density matrix $\rho $ satisfies the  quantum-mechanical optical 
Bloch equation
\be
i\hbar ( {\pp \over \pp t } + v{\pp \over \pp x})\rho = [(H_{0} -
{\bf E}\cdot {\bf d}) \ , \ \rho  ] \ .
\label{Bloch}
\ee
$H_{0}$ denotes the Hamiltonian of a free atom and $v$ is the 
$x$-component of the velocity of the atoms. In most cases, SVEA 
is further made which assumes that the amplitudes 
$ {\cal {\bf E}}_{l} $ vary slowly compared to the space and time 
scales determined by $k_{l}$ and $w_{l}$. Under SVEA, the 
Maxwell-Bloch equation becomes a set of coupled first order partial 
differential equations for the amplitudes $ {\cal {\bf E}}_{l} $ and 
the components of the density matrix. Explicit expressions of the 
Maxwell-Bloch equation for some physically relevant cases are given 
in Sec. 3. Thus, instead of using quantum electrodynamics for the 
interaction of radiation with matter, a specific choice of physical 
systems allows an effective description using the amplitudes 
$ {\cal {\bf E}}_{l} $ as a reduced set of variables.  This makes the 
use of potential variables quite difficult if not impossible. Also, 
lack of potential variables causes the physical system to be described 
only by the equation of motion, not by the action principle. 
Consequently, a field theoretic formulation is lacking in pulse 
propagation problems. However, there exists one exceptional case. 
When pulses propagate in a resonant, nondegenerate two-level 
atomic medium with inhomogeneous broadening, McCall and Hahn have 
introduced an effective potential-like variable and shown 
that an arbitrary pulse evolves into a coherent mode of 
lossless pulses. This phenomenon, known as self-induced transparency, 
is also observed in more general, degenerate and/or multi-level cases. 
Specifically, McCall and Hahn have shown that when the pulse envelope 
is assumed to be real and the time area of the dimensionless envelope 
function $2E$ ,
\be
\theta (x) = \int_{-\infty }^{ \infty } dt 2E ,
\label{area}
\ee
is an integer multiple of $2\pi $ ( $2n\pi $ pulse), then the pulse 
propagates without loss of energy. Otherwise, due to inhomogeneous 
broadening the pulse quickly reshapes to $2n \pi $ pulse  according to 
the area theorem,
\be
 {d \theta (x) \over dx } = -\a \sin{ \theta (x) } ,
\label{areathm}
\ee
for some constant $\a $.
These $2n \pi $ pulses are another type of optical solitons which arise 
from the nonlinear response of matter to the radiation field besides the 
most well-known optical solitons of the nonlinear Schr\"{o}dinger 
equation appearing in optical communication problem. The proof of the 
area theorem follows from inhomogeneous broadening as well as the 
Maxwell-Bloch equation. The SIT equation for the nondegenerate 
two-level case is given in a dimensionless form by 
\ben
\pb E + 2 \b <P> &=& 0 \nonumber \\
\pp D - E^{*}P - EP^{*} &=& 0 \nonumber \\
\pp P + 2i\xi  P + 2ED &=& 0
\label{sit}
\een
where $\b $ is an arbitrary constant and 
$\xi  = w-w_{0} \ , \ \pp \equiv \pp /\pp z \ , \ \pb \equiv \pp / 
\pp \bar{z} \ , z= t-x/c , \bar{z} = x/c$. The angular bracket 
signifies an average over the spectrum $f(\xi  )$ as given by
\be
< \cdots > = \int^{\infty }_{- \infty } ( \cdots )f(\xi  )d\xi .
\label{inhomog}
\ee
The dimensionless quantities 
$E,P$ and $ D$ correspond to the electric field, the polarization and 
the population inversion through the relation,
\ben
E &=& -i{\bf E}\cdot{\bf e} t_0 d/\sqrt{6} \hbar \nonumber \\ 
P &=& -\r_{12} \exp [-i(kx-\o t)]/4k t_0 N_0 f(\x ) \nonumber \\ 
D &=& -(\r_{22}-\r_{11})/8 k t_0 N_0 f(\x ) 
\label{epd}
\een
where ${\bf e}$ specifies the linear polarization direction, 
$t_0$ is a time constant and $N_0$ is related to the stationary 
populations of the levels.\footnote{For the details of constants, we 
refer the reader to ref. \cite{maim2}.} 
In order to understand the structure of $2n \pi $ pulses 
better, we may impose further restrictions such that the system is on 
resonance ($\xi  =0$) without frequency modulation ($E$ being real) and 
inhomogeneous broadening ($f(\xi ) = \s (\xi )$). Under such 
circumstances, we can introduce an area function $\vf (x,t)$,
\be
\vf (x,t) = \int_{-\infty }^{t} Edt^{'}
\label{areaftn}
\ee
which in the limit $t \rightarrow \infty $ agrees with $\theta(x)/2$ 
in (\ref{area}). In terms of $\vf $, the SIT equation reduces to the 
sine-Gordon equation,
\be
\pb\pp \vf - 2\b \sin{2\vf } = 0 ,
\label{sg}
\ee
together with the identification;
\be
E=E^{*} = \pp \vf \ \ , \  \ <P> = P = -\sin{2\vf } \ \ , \ \ 
<D> = D= \cos{2\vf } \ .
\label{epd2}
\ee
The sine-Gordon equation arises from the Lagrangian
\be
S =  {1 \over 2\pi } \int ( \pp \vf \pb \vf - 2\b \cos{2\vf } ) .
\ee
The periodic cosine potential term exhibits infinitely many degenerate 
vacua and gives rise to soliton solutions which interpolate between 
different vacua. This shows that $2n \pi $ pulse is in fact the 
topological $n$-soliton solution of the sine-Gordon equation in this 
particular limit. The electric field amplitude $E$, now identified with 
$\pp \vf $, receives an interpretation as a topological current. 
Note that the effective potential variable $\vf $ is different from the 
conventional scalar or vector potentials of the electromagnetic field. 
Nevertheless, it is remarkable that the potential energy $\cos {2\vf }$ 
of the sine-Gordon Lagrangian can be identified with the population 
inversion $D$ which represents the atomic energy. Also the Lorentz 
invariance, which was broken by SVEA, re-emerges in the 
sine-Gordon field theory after the redefinition of coordinates. 
The identification of the atomic energy with the cosine potential term 
possessing degenerate vacua, consequently topological solitons, 
underlies the stability of $2n\pi $ pulses.

Unfortunately, the sine-Gordon theory formulation of SIT is too 
restrictive. In the presense of frequency modulation, $E$ should be 
complex. Therefore, it can not be replaced by a real scalar field 
$\vf $ and the sine-Gordon limit is no longer valid. Also, frequency 
modulation effects invalidate the area theorem. By a direct application 
of the inverse scattering, it has been found, however, that solitons 
do exist even in the case of complex $E$ \cite{lamb}. This suggests 
that a more general field theory of SIT than the sine-Gordon theory 
could exist in order to account for complex $E$. Recently, we have 
shown that this is indeed true and the generalizing theory which 
includes both frequency detuning and modulation effects is given by the 
so-called complex sine-Gordon theory in the following way \cite{ps1}; 
assume that $E$ is complex and the frequency distribution function of 
inhomogeneous broadening is sharply peaked at $\xi $, 
i.e. $f(\xi ^{'}) = \d (\xi ^{'}  - \xi )$ for some constant $\xi $.  
Introduce parametrizations of $E, ~P$ and $D$ which generalize 
(\ref{epd2}) in terms of three scalar fields $\vf , ~ \q $ and $\h $,
\be
E = e^{i(\q - 2\h )}( 2\pp \h {\cos{\vf } \over \sin{\vf }} - 
i\pp \vf )  \ \ , \ \ 
P = ie^{i(\q - 2\h )}\sin{2\vf } 
\ \ , \ \ 
D = \cos{ 2\vf } \ .
\label{csgepd}
\ee
These parametrizations consistently change the SIT equation (\ref{sit}) 
into a couple of second order nonlinear differential equations known as 
the complex sine-Gordon equation;
\be
\pb\pp \vf + 4{\cos{\vf } \over \sin^{3}{\vf }}\pp \h \pb \h -
2\b \sin{2\vf }  = 0 
\label{cs}
\ee
\be
\pb \pp \h - {2 \over \sin{2\vf }}(\pb \h \pp \vf + \pp \h \pb \vf 
) = 0
\label{csg}
\ee
and a couple of first order constraint equations,
\ben
2\cos^{2}{\vf }\pp \h - \sin^{2}{\vf }\pp \q - 2\xi \sin^{2}{\vf } 
&= & 0 \nonumber \\
2\cos^{2}{\vf }\pb \h + \sin^{2}{\vf }\pb \q  &=& 0 \ .
\label{constraint}
\een
Note that the complex sine-Gordon equation reduces to 
the sine-Gordon equation when frequency modulation effect is ignored 
such that $\h =0, \ \q = \pi / 2 $ and the system is on 
resonance ($\xi = 0$). This reduction is consistent with the original 
equation since solutions of the sine-Gordon equation consists a 
subspace of the whole solution space. The complex sine-Gordon equation 
was first introduced by Lund and Regge in 1976 in order to describe 
the motion of relativistic vortices in a superfluid  \cite{lund}, 
and also independently by Pohlmeyer in a reduction problem of O(4) 
nonlinear sigma model \cite{pohl}. The integrability and various 
related properties, both classical and quantum, of the complex 
sine-Gordon equation have been studied since then. In particular, the 
Lagrangian for the complex sine-Gordon equation in terms of $\vf $ and 
$\h $ is given by
\be
S = {1 \over 2\pi }\int \pp \vf \pb \vf + 4 \cot^{2}{\vf } \pp \h \pb \h 
- 2\b \cos {2\vf } .
\label{csglag}
\ee
This Lagrangian, however, is singular at $\vf = n \pi $ for integer $n$ 
which causes difficulties in quantizing the theory. Also, besides the 
complex sine-Gordon equation, the SIT equation comprises the constraint 
equation (\ref{constraint}). Thus the Lagrangian (\ref{csglag}) does not 
quite serve for a field theory of two-level SIT. In fact, the singular 
behavior of the Lagrangian (\ref{csglag}) is an artifact of neglecting 
the constraint equation. This fact as well as the rationale of the above 
parametrizations can be seen most clearly if we reformulate 
the Lagrangian to include the constraint in the context of a nonlinear 
sigma model as explained in the next section.

\section{Effective field theory }
\setcounter{equation}{0}

In order to construct a field theory of SIT in terms of potential 
variables and also find a way to extend to more general multi-level 
and degenerate cases, we first note that the optical Bloch equation 
admits an interpretation as a spinning top equation like the 
corresponding magnetic resonance equations \cite{bloch}. Denote real 
and imaginary parts of $E$ and $P$ by $E = E_{R} + iE_{I}, ~ P = 
P_{R} + iP_{I}$. Then, the Bloch part of the SIT equation (\ref{sit}) 
can be expressed as 
\be
\pp \vec{S} = \vec{\Omega } \times \vec{S}
\label{topeq}
\ee
where $\vec{S} = (P_{R}, ~ P_{I}, ~ D), ~~ \vec{\Omega } = 
(2E_{I}, ~ -2E_{R}, ~-2\xi )$, i.e. it describes a spinning top 
where the electric dipole ``pseudospin" vector $\vec{S}$ precesses 
about the ``torque" vector $\vec{\Omega }$. This clearly shows that 
the length of the vector $\vec{S}$ is conserved,
\be
|\vec{S}|^{2} = P_{R}^{2} + P_{I}^{2} + D^{2} = 1,
\label{prob}
\ee
where the length equals unity due to the conservation of probability.
The remaining Maxwell part of the SIT equation determines strength of 
the torque vector. If $P_{I} = 0$, we may solve (\ref{prob}) by taking 
$P_{R} = -\sin{2\vf } $ and $D = \cos{2\vf }$ and also (\ref{topeq}) 
by taking $E = \pp \vf $ as given in (\ref{epd2}). Then, the Maxwell 
equation becomes the sine-Gordon equation as before. This picture 
agrees with the conventional interpretation of the sine-Gordon theory 
as describing a system of an infinite chain of pendulums.
In order to generalize the sine-Gordon limit to the complex $E$ and 
$P$ case, we make a crucial observation that the constraint (\ref{prob}) 
in general can be solved in terms of an $SU(2)$ matrix potential 
variable $g$ by
\be
\pmatrix{D & P \cr P^{*} & -D} = g^{-1}\s_{3}g, ~~~ \s_{3} = 
\pmatrix{1 & 0 \cr 0 & -1} .
\label{gspin}
\ee
The identity equation,
\be
\pmatrix{ \pp D & \pp P \cr \pp P^{*} & -\pp D } 
=\pp (g^{-1} \s_{3} g) = [ g^{-1}\s_{3}g, ~ g^{-1} \pp g],
\ee
gives the Bloch equation if we make an identification,
\be
g^{-1} \pp g -R =  \pmatrix{ i\xi  & -E \cr E^{*} & -i\xi } ,
\label{idelec}
\ee 
where $R$ is an arbitrary matrix commuting with $g^{-1}\s_{3}g$. 
The identification (\ref{gspin}) requires that $\mbox{Tr} g^{-1}
\s_{3} g =0$ and $(g^{-1} \s_{3} g)^T = (g^{-1} \s_{3} g)^*$. 
In other words, $g$ is a unitary matrix. Similarly, the tracelessness 
of $g^{-1} \pp g$ from the identification (\ref{idelec}) further 
requires $g$ to be an $SU(2)$ matrix. We will show shortly that $R$ is 
determined by demanding a Lagrangian for the above equation and is 
also traceless. Finally, the Maxwell equation becomes
\ben
\pb (g^{-1} \pp g -R) &=& \pmatrix{ 0 & -\pb E \cr \pb E^{*} & 0 } 
= - \left[ \  \pmatrix{ i\b  & 0 \cr 0 & -i\b  } \ , 
~  i \pmatrix{D & P \cr P^{*} & -D }
\right] \nonumber \\ 
&=&  \b[ \s_{3}, ~ g^{-1}\s_{3} g].
\label{maxpot}
\een
Thus, we have successfully expressed the SIT equation in terms of 
potential variable $g$ up to an undetermined quantity $R$. Note that 
the diagonal part of the r.h.s. of (\ref{idelec}) is fixed to a 
constant $i\xi $ accounting for frequency detuning. This constrains 
$g$ such that the diagonal part of $ g^{-1} \pp g -R $ is equal to  
$(i\xi ,~ -i\xi )$. Therefore, for a field theory of SIT, we shall 
construct a Lagrangian in terms of the potential variable $g$ whose 
equation of motion is (\ref{maxpot}) together with a Lagrange multiplier 
for the constraint. In order to help understanding, we assume for a 
moment that $R = 0$ and the system is on resonance ($\xi = 0$). Then, 
the equation of motion (\ref{maxpot}) arises from a variation of the 
action
\be
S = S_{WZW}(g)  - S_{\mbox{pot}} + S_{\mbox{const}}
\label{act}
\ee 
with the following variational behaviors;
\be
\d _{g}S_{WZW} =  {1 \over 2\pi }\int \mbox{Tr}
\pb ( \gi \pp g ) \gi \d g , ~~~ 
\d_{g}S_{\mbox{pot}} = {\b \over 2\pi }\int \mbox{Tr}
 [ \s_{3}, ~ \gi \s_{3} g \ ]\gi \d g .
\label{zeroeq}
\ee
The action $S_{WZW}(g)$ is the well-known $SU(2)$ Wess-Zumino-Witten 
functional,
\be
S_{WZW}(g)=-{1\over 4\pi }\int_{\S }\mbox{Tr } \gi \pp g \gi \pb g 
- {1 \over 12\pi }\int_{B}\mbox{Tr } \tilde{g}^{-1}d \tilde{g}\wedge 
\tilde{g}^{-1}d \tilde{g} \wedge \tilde{g}^{-1}d \tilde{g} \ ,
\label{wzw}
\ee
where the second term on the r.h.s., known as the 
Wess-Zumino term, is defined on a three-dimensional manifold $B$ with 
boundary $\S $ and $\tilde{g}$ is an extension of a map $g:\S 
\rightarrow SU(2)$ to $B$ with $ \tilde{g}|_{ \S }=g$ \cite{Witten}.
The potential term $S_{\mbox{pot}}$ can be easily written by
\be
S_{\mbox{pot}}= {\b \over 2\pi }\int \mbox{Tr}g\s_{3} \gi \s_{3} .
\label{potential}
\ee
Finally, the constraint requires vanishing of the diagonal part of the 
matrix $g^{-1}\pp g$ which can be imposed by adding a Lagrange 
multiplying term $S_{\mbox{const}}$ to the action
\be
S_{\mbox{const}} = {1 \over 2\pi }\int \mbox{Tr} \L \s_{3} g^{-1}\pp g .
\ee
The Lagrange multiplier $\L $ however adds a new term to the equation 
of motion by
\be
\d_{g}S_{\mbox{const}} = {1 \over 2\pi }\int \mbox{Tr}(-\pp \L\s_{3} + 
[\L \s_{3}, ~ g^{-1}\pp g ])g^{-1}\d g ,
\ee
which seems to spoil our construction of a SIT field theory. 
This problem can be resolved beautifully if we lift the action 
(\ref{act}) to the ``vector gauge invariant" one by replacing the 
constraint term with a ``gauging" part of the Wess-Zumino-Witten 
action,
\ben
S &=& S_{WZW}(g)  - S_{\mbox{pot}} + S_{\mbox{gauge}} 
\label{action} \\
S_{\mbox{gauge}} &=& 
{1 \over 2\pi }\int \mbox {Tr} (- A\pb g \gi + \Ab \gi \pp g
 + Ag\Ab \gi - A\Ab )
\een
where the connection fields $A, \Ab$ gauge the anomaly free subgroup 
$U(1)$ of $SU(2)$ generated by the Pauli matrix $\s_{3}$. They 
introduce a vector gauge invariance of the action under the transform
\be
g \rightarrow h^{-1}gh \ \ , \ \ A \rightarrow h^{-1}Ah + 
h^{-1}\pp h \ \ , \ \  \Ab \rightarrow h^{-1}\Ab h  + h^{-1}\pb h 
\label{gaugetr}
\ee
where $h:\S \rightarrow U(1)$. Owing to the absence of kinetic terms, 
$A, \Ab $ act as Lagrange multipliers which result in the constraint 
equations when integrated out. The action (\ref{act}) may be understood 
as a particular gauge fixing of the vector gauge invariance where 
$A = 0, ~ \Ab = \L \s_{3}$. However, a more convenient gauge fixing 
which manifests the equivalence with the SIT equation (\ref{sit}) is 
where $A= i\xi \s_{3} , ~ \Ab =0$. Such a gauge fixing is always 
possible as a result of (\ref{flat}). Before proving this, we discuss 
about the generalization of the action (\ref{action}) to groups other 
than $SU(2) \supset U(1) $. We may simply replace the pair $SU(2) 
\supset U(1) $ by $G \supset H $ for any Lie groups $G$ and $H$ and 
obtain the $G/H$ gauged Wess-Zumino-Witten action ($S_{WZW} + 
S_{\mbox{gauge}}$). This action is known to possess conformal 
symmetry and has been identified with the action for the general 
$G/H$ coset conformal field theory \cite{coset}. The potential term 
(\ref{potential}) breaks conformal symmetry. Nevertheless, it preserves 
the integrability of the model given by (\ref{action}) where $G/H = 
SU(2)/U(1)$ and this model has been used in describing integrable 
perturbation of certain coset conformal field theories \cite{bakas,park}. 
For a general pair of $G$ and $H$, the expression for the potential 
which preserves integrability has been also found \cite{park,shin2}. 
It is given by
\be
S_{\mbox{pot}}= {\b \over 2\pi }\int \mbox{Tr}gT\gi \Tb
\label{potent}
\ee
where $T$ and $ \Tb $ are constant matrices which commute with the 
subgroup $H$, i.e. $[T, h] = [\Tb , h ] = 0, \mbox{ for } h \in H$ so 
that the potential term is vector gauge invariant. In general, 
$S_{\mbox{pot}}$ is specified algebraically by a triplet of Lie 
groups $F \supset G \supset H$ for every symmetric space $F/G$, where 
the Lie algebra decomposition ${\bf f} = {\bf g} \oplus {\bf k} $ 
satisfies the commutation relations,
\be
[{\bf g} ~ , ~ {\bf g}] \subset {\bf g} ~ , ~ [ {\bf g} ~ , ~ {\bf k}] 
\subset {\bf k} ~ , ~ [{\bf k} ~ , ~ {\bf k} ] \subset {\bf g} ~ .
\label{algebra}
\ee
We take $T$ and $\Tb$ as elements of $ {\bf k}$ and  define  
$ {\bf h} $ as the simultaneous centralizer of $T$ and $\Tb $, i.e. 
${\bf h} = C_{{\bf g}}(T, \Tb ) = \{ B \in {\bf g} \ : \ [B ~ , ~ T] = 
0 = [B ~ , ~ \Tb]\} $ with $H$ its associated Lie group. 
With these specifications, the action (\ref{action}) becomes integrable 
and generalizes the sine-Gordon model according to each symmetric 
spaces \cite{bps}. For compact symmetric spaces of type II, e.g. 
symmetric spaces of the form $G \times G/G$, the elements $g$ and 
$T$ take the form $g \otimes g$ and $T \otimes 1 - 1 \otimes T$ 
(and similarly for $\Tb $). In which case, the model becomes 
effectively equivalent to the case where $T, ~ \Tb$ belong to the 
Lie algebra ${\bf g}$. Thus the model is specified by the coset $G/H$ 
where $H$ is the stability subgroup of $T$ and $\Tb $ for $T, ~ \Tb 
\in {\bf g }$. In this paper, we will restrict ourselves only to 
this case. As we will see later, physically interesting cases of SIT 
all correspond to even more specific symmetric spaces, where $G/H$ 
becomes Hermitian symmetric spaces and the adjoint action of $T$ 
defines a complex structure on $G/H$.

Now, we demonstrate the integrability of the model by expressing the 
equation of motion arising from the action (\ref{action}) in a zero 
curvature form,
\ben
\d_{g}S  &=& -{1 \over 2\pi }\int \mbox{Tr}
([\  \pp + \gi \pp g + \gi A g  , ~ \pb + \Ab ] + \b [T, ~ \gi \Tb g ])
\gi \d g  \nonumber \\
&=& -{1 \over 2\pi }\int \mbox{Tr}
[\  \pp + \gi \pp g + \gi A g + \b\l T \ , \ \pb + \Ab + 
{1 \over \l }\gi \Tb g \ ]\gi \d g \nonumber \\ 
&=& 0
\label{zeroeqn}
\een
where $\l $ is an arbitrary complex constant. That is, the equation of 
motion is given by a zero curvature condition in terms of a $U-V$ pair, 
\be
[\pp - U, ~ \pb -V]=0,
\label{uvpair}
\ee
where
\be
U \equiv - \gi \pp g - \gi A g - \b\l T, ~~ 
V \equiv - \Ab - {1 \over \l }\gi \Tb g .
\label{uv}
\ee
This shows that the equation of motion arises as the integrability of 
the overdetermined linear equations;
\be
(\pp + \gi \pp g + \gi A g + \b\l T  ) \Psi = 0 , ~~~
( \pb + \Ab + {1 \over \l } \gi \Tb g  ) \Psi = 0 .
\label{lineareqn}
\ee
The constraint equations coming from the $A, \Ab $-variations are
\ben
\d _{A}S &=& {1 \over 2\pi }\int \mbox{Tr} ( \ - \pb g 
\gi + g\Ab \gi - \Ab \  )\d A
= 0  \nonumber \\
\d _{\Ab }S &=& {1 \over 2\pi }\int \mbox{Tr} ( \  \gi 
\pp g  +\gi A g - A \ )\d\Ab = 0 \ .
\label{constraint2}
\een
Note that these constraint equations when combined with (\ref{zeroeqn}) 
imply the flatness of the connection $A$ and $ \Ab $, i.e.
\be
F_{z \zb } = [ \ \pp + A \ , \ \pb + \Ab \ ] = 0 \ .
\label{flat}
\ee
In the following, we show that (\ref{zeroeqn}) and (\ref{constraint2}) 
can be identified with the SIT equation for various cases depending on 
the choices of the groups $G$ and $H$,  and $T , \Tb $ as well as the 
specific choice of gauge fixing. Thus, we obtain the action principle 
of SIT equations when inhomogeneous broadening is ignored. The inclusion 
of inhomogeneous broadening is also obtained in our field theoretic 
context. Remarkably, in the presense of inhomogeneous broadening, the 
notion of effective potential still persists and the inhomogeneous 
broadening effect, i.e. Doppler shifted atomic motions, can be 
beautifully incorporated into the $U(1)$ vector gauge transformation 
of the theory. 
\subsection{Examples}
\vglue .2in
{\bf Nondegenerate two-level system}
\vglue .2in
This is the simplest case of SIT which was originally considered by 
McCall and Hahn. It also accounts for the transitions $1/2 \rightarrow 
1/2, ~ 1\leftrightarrow 0, ~ 1 \rightarrow 1 $ and $3/2 \leftrightarrow 
1/2$ for linearly polarized waves and the transitions $1/2 \rightarrow 
1/2, ~ 1 \leftrightarrow 0 $ and $ 1 \rightarrow 1$ for circularly 
polarized waves. The SIT equation is given by (\ref{sit}) which can 
be expressed in an equivalent zero cuvature form,
\be
\left[ \  \pp + \pmatrix{ i\b \l + i\xi & -E \cr E^{*} & -i\b 
\l -i\xi } \ , \ \pb - {i \over \l } \pmatrix{D & P \cr P^{*} & -D }
\right]  = 0 \ .
\label{sitzero}
\ee
In order to show that this SIT equation in fact arises from the 
effective field theory (\ref{action}), we take $H= U(1) \subset  
SU(2) = G$ and $ T = - \Tb = i\s _{3} = \mbox{diag}( i , -i )$ for 
Pauli matrices $\s _{i}$. We fix the vector gauge invariance by choosing 
\be 
A = i\xi \s_{3} \ , \ \Ab =0
\label{gfix}
\ee
for a constant $\xi $. Such a gauge fixing is possible due to the 
flatness of $A, \Ab$. Comparing (\ref{uv}) with (\ref{sitzero}), we 
could identify  $E, P$ and $D$ in terms of $g$ such that
\be
g^{-1}\pp g + \xi  g^{-1}Tg - \xi  T = \pmatrix{ 0 & -E  \cr E^{*} 
& 0 } \ \ , \ \ g^{-1}\bar{T}g = -i \pmatrix{ D & P \cr P^{*} & -D }
\label{epdtwo}
\ee
which are consistent with the constraint equation (\ref{constraint2}). 
Comparing with (\ref{gspin})-(\ref{idelec}), we find that $R$ is fixed 
to $-i\xi g^{-1}\s_{3} g$. 
Then, the zero curvature equation (\ref{zeroeqn}) agrees precisely with 
(\ref{sitzero}). Furthermore, if we make an explicit parametrization 
of the $2\times 2 ~ SU(2)$ matrix $g$ by
\be
g=e^{i\eta \s_{3}}e^{i\varphi (\cos{\q }\s_{1} -\sin{\q }\s_{2})}e^{i\eta 
\s_{3}}= \pmatrix{ e^{2i\eta }\cos{\varphi } & i\sin{\varphi }e^{i\q } 
\cr i\sin{\varphi }e^{-i\q } & e^{-2i\eta }\cos{\varphi } }\ ,
\label{para}
\ee
we recover the parametrizations of $E , P $ and $D$ as given in 
(\ref{csgepd}) and the SIT equation becomes the complex sine-Gordon 
equation (\ref{cs})(\ref{csg}) and the constraint equation 
(\ref{constraint}). The potential term in (\ref{action}) now changes 
into the population inversion $D$,
\be
S_{\mbox{pot}} = \int {\b \over \pi }\cos{2\varphi } = 
\int {\b \over \pi }D   ,
\label{twopot}
\ee
which for $\b >0 $ possesses degenerate vacua at
\be
\varphi = \vf_{n} = (n+ {1\over 2} )\pi , \ n \in Z  \ \mbox{ and } \  
\q \ = \q_{0} \ \ \mbox{for} \ \ \q_{0} \  \mbox{constant} \ .
\label{twovac}
\ee
The property of degenerate vacua and the corresponding soliton solutions 
will be considered in Sec. 4.
\vglue .2in
{\bf Degenerate two-level system}
\vglue .2in
One of the deficiencies of the SIT model of McCall and Hahn is the 
absence of level degeneracy. Since most atomic systems possess level 
degeneracy, the analysis of the nondegenerate two-level system does not 
apply to a more practical system. Moreover, level degeneracy in general 
breaks the integrability and does not allow exact soliton 
configurations. For example, propagation of pulses in a two-level 
medium with the transition $j_{b} = 2 \rightarrow j_{a} = 2$ 
is effectively described by the double sine-Gordon equation
\be
\pp\pb \vf = c_{1} \sin {\vf } + c_{2} \sin{2 \vf }
\label{dsg}
\ee
which is not integrable. Nevertheless, there are a few exceptional 
cases which are completely integrable even in the presense of level 
degeneracy. It was shown that \cite{maim,bash1} the SIT equations for 
the transitions $j_{b} = 0 \rightarrow j_{a} = 1 , \ j_{b} = 1 
\rightarrow j_{a} = 0 $ and $j_{b} =1 \rightarrow j_{a} = 1 $ are 
integrable in the sense that the SIT equations can be expressed in an 
inverse scattering form. In the following, we show that these cases 
correspond to the effective theory with $G = SU(3) $ and $ H = U(2) 
\subset G $. Also, we show that the local vector gauge structure 
incorporates naturally the effects of frequency detuning and 
longitudinally applied magnetic field. Consider a monochromatic pulse 
propagating through a medium of degenerate two-level atoms in the 
presence of a longitudinal magnetic field. Then, the Maxwell-Bloch 
equation under SVEA is given by 
\begin{eqnarray*}
\pb \ve ^{q} = i \sum_{\m m} \langle R_{\m m} \rangle J^{q}_{\m m} 
\end{eqnarray*}
\begin{eqnarray*}
[\pp + i ( 2 \xi  + \O_ {b} \m - \O_ {a} m )] R_{\m m } =
 i \sum_{q} \ve ^{q}( \sum_{m^{'}} J^{q}_{\m m^{'}} R_{m^{'} m} - 
 \sum_{\m ^{'}} R_{\m \m ^{'}} J^{q}_{\m ^{'} m}   )
\end{eqnarray*}
\begin{eqnarray*}
[\pp + i\O_ {a}(m - m^{'}) ] R_{m m^{'}} =
i \sum_{q \m } (\ve ^{q *} J^{q}_{\m m } R_{\m  m^{'} }  - 
 \ve ^{q } J^{q}_{\m m^{'} } R_{ m \m  } )
\end{eqnarray*}
\be
[\pp + i\O_ {b}(\m - \m^{'})] R_{\m \m^{'}} = 
i \sum_{q m } (\ve ^{q } J^{q}_{\m m } R_{m  \m^{'} }  - 
 \ve ^{q *} J^{q}_{\m^{'} m } R_{ \m m  } ) .
\label{mb1}
\ee
The dimensionless quantities ${\bf \ve }^q $ and $ { R }$ are 
propotional to the electric field amplitude ${ E}$ 
and the density matrix ${\bf \r }$, 
where $q$ is the polarization index and the subscripts $\m , 
\m^{'} , \dots $ and $ m , m^{'} , \dots $ denotes projections of the 
angular momentum on the quantization axis in two-level states $ |a > $ 
and $ |b > $ respectively.\footnote{For details of propotionality 
constants and their physical meanings, we refer the reader to ref.  
\cite{maim2}.}  $J$ denotes the Wigner's $3j$ symbols
\be
J^{q}_{\m m} = (-1)^{j_{b}-m} \sqrt{3} \pmatrix{ j_{a} & 1 & j_{b} \cr
-m & q & \m },
\label{wigner}
\ee
and $\O _{a} (\O _{b} )$ is a dimensionless coupling constant of an 
external magnetic field.

In general, (\ref{mb1}) is not integrable. However, with particular 
choices $j_{a}$ and $j_{b}$, (\ref{mb1}) can be recasted into the zero 
curvature form, or the $U-V$ pair as in (\ref{uvpair}).
Specifically, for the transition $j_{b} = 1/2 \rightarrow j_{a} = 1/2$, 
$U$ and $ V$ becomes
\be
U = \pmatrix{ U_{+} & 0 \cr 0 & U_{-} } ~ , ~ V = \pmatrix{ V_{+} & 0 
\cr 0 & V_{-} } 
\ee
where
\ben
U_{\pm } &=& \pmatrix{ -i(x + \l ) & \pm i\ve ^{\pm 1} \cr \mp i 
\ve^{\pm 1 * } & i(x + \l ) } 
~ , ~ V_{\pm } = -{1\over 2\l }\pmatrix{R^{(b)}_{\mp{1\over 2} 
\mp{1\over 2} } &  R^{(ba)}_{\mp{1\over 2} \pm{1\over 2} } 
\cr R^{(ba)*}_{\mp{1\over 2} \pm{1\over 2} } & 
R^{(a)}_{\pm{1\over 2} \pm{1\over 2} } } \nonumber \\
x &=& {1\over 4}(\O _{a} + \O _{b} - 4\xi  ) .
\label{uvhalf}
\een
In the context of effective field theory, we again identify 
the $U-V$ pair in terms of $g$ by
\be
U = - \gi \pp g - \gi A g - \b\l T, ~~~~ V = - {1 \over \l }\gi \Tb g 
\ee
where the gauge choice is
\be
A= \pmatrix{-ix & 0 & 0 & 0 \cr
		  0 & ix & 0 & 0 \cr
		 0 & 0 & -ix & 0 \cr
		 0 & 0 & 0 & ix }  , ~~ \Ab = 0,
\ee
and
\be
T = -{\bar T} = {i} \pmatrix{ \s_3 & 0 \cr 0 & \s_3 }
\ee
with the Pauli matrix $\s_3$. Here we set $\b =1$ for convenience.
The resulting effective theory is specified 
by  the coset $G/H = (SU(2) \times SU(2))/ (U(1) \times U(1))$
such that $g = \pmatrix{ g_1 & 0 \cr 0 & g_2 }$ with $g_1, g_2 
\subset SU(2)$ and the two $U(1)$ subgroups are generated by 
$\pmatrix{ \s_{3} & 0 \cr 0 & 0 } $ and $\pmatrix{ 0 & 0 \cr 0 & 
\s_{3} }$.  Note that the specific form
of the identification (\ref{uvhalf}) requires $g_1$ and $g_2$ to be
$SU(2)$ as in the case of the nondegenerate two-level system. Thus, 
this case is simply two sets of the nondegenerate two-level case.

Another integrable cases are the transitions; $j_{b} = 1 \rightarrow 
j_{a} = 0 $ and $ j_{b} = 0 \rightarrow j_{a} = 1 $.
In each cases, the $U-V$ pair is given by
\ben
U &=& \pmatrix{ -{4\over 3}i\l + i(x+y)& -i\ve^{-1} & -i\ve^{1} \cr
		 -i\ve^{-1*} & {2\over 3}i\l -ix & 0 \cr
		 -i\ve^{1*} & 0 & {2\over 3}i\l -iy }
\nonumber \\
&& \nonumber \\
x&=& -\O_{a} - {2\over 3}\xi  , ~~ y = \O_{a} - {2\over 3}\xi  ~ ~ ~ 
\mbox{for} ~ j_{b} = 0 \rightarrow j_{a} = 1 \nonumber \\
x&=& -\O_{b} - {2\over 3}\xi  , ~~ y = \O_{b} - {2\over 3}\xi  ~ ~ ~ 
\mbox{for} ~ j_{b} = 1 \rightarrow j_{a} = 0 .
\label{Uonly}
\een
and
\ben
V &=& {i \over 2\l }\pmatrix{ R^{(b)}_{00} & R^{(ba)}_{0-1} & 
R^{(ba)}_{01} \cr R^{(ba)*}_{0-1}& R^{(a)}_{-1-1} & 
R^{(a)}_{-11} \cr R^{(ba)*}_{01} & R^{(a)}_{1-1} & R^{(a)}_{11}}
~ \mbox{ for } ~ j_{b} = 0 \rightarrow j_{a} = 1 \nonumber \\
&& \nonumber \\
V&=& {i \over 2\l }\pmatrix{ -R^{(a)}_{00} & R^{(ba)}_{10} & 
R^{(ba)}_{-10} \cr R^{(ba)*}_{10}& -R^{(b)}_{11} & -R^{(b)}_{-11} 
\cr R^{(ba)*}_{-10} & -R^{(b)}_{1-1} & -R^{(b)}_{-1-1}}
~ \mbox{ for } ~ j_{b} = 1 \rightarrow j_{a} = 0.  
\een
The gauge fixing is given by
\be
A= \pmatrix{-i(x+y)  & 0 & 0 \cr
		  0 & ix  & 0 \cr
		 0 & 0 & iy }, ~~ \Ab = 0 ,
\ee
and the effective field theory is specified by $ G/H = SU(3)/U(2)$ with
\be
T = -{\bar T} = {2 i \over 3} \pmatrix {2 & 0 & 0 \cr 0 & -1 & 0 \cr
0 & 0 & -1 }
\ee
and $\b =1$. It is interesting to observe that the electric field 
components $\ve^{1} , \ve^{-1}$ in (\ref{Uonly}) parametrize the coset 
$SU(3)/U(2)$ and the vector $(\ve^{-1*} , \ve^{1*})^{T}$ transforms as 
a vector under the $U(2)$ action. In particular, since frequency 
detuning amounts to the global $U(1) ( \subset U(2)) $ action while 
longitudinal magnetic field amounts to the global $U(1) \times U(1) 
(\subset U(2))$ action, the effects of detuning and magnetic field to 
$\ve^{1} , \ve^{-1}$ can be easily obtained.
\vglue .2in
{\bf Three level system}
\vglue .2in
The propagation of pulses in a multi-level medium
with several carrier frequencies as given in (\ref{elec})   
is a more complex problem than the two-level SIT and in general is 
not an integrable system. However, with certain restrictions on the 
parameters of the medium, it becomes integrable again and reveals 
much richer structures. Here, we will not exhaust all integrable 
multi-levels cases but restrict only to the degenerate three-level case 
and provide an effective field theory formulation. Other multi-level 
cases can be treated in a similar way. The propagation of two-frequency 
light  interacting with a three-level medium, having either the $\L $ 
or $V$-type resonance configurations, has been studied earlier 
and its integrability has been 
demonstrated in the context of the inverse scattering 
method \cite{Bash2}. We suppress the general Maxwell-Bloch equation 
formulation for the three-level case and refer the reader to 
Ref. \cite{Bash2} for details. Here, we extend the Maxwell-Bloch 
equations of Ref. \cite{Bash2} in order to include a longitudinal 
magnetic field. Then, the Maxwell-Bloch equation in a dimensionless 
form, describing the $\L $ configuration with $j_{b} = 1, j_{a} = 
j_{c} = 0 $, is given by
\be
\pb \ve^{q}_{j} = -i p^{q}_{j}, ~~~ j = 1,2, ~~~ q = \pm 1
\ee
and
\ben
( \pp + i t_{0}(k_{1}v - 2 \D _{1} - \O _{b} q)) p^{q}_{1} &=& 
-i(\sum_{q^{'}} \ve_{1}^{q^{'}}m_{q^{'}q} - \ve_{1}^{q} n_{1} - 
\ve^{q}_{2}r ) \nonumber \\ ( \pp + i t_{0}(k_{2}v - 2 \D _{2} - 
\O _{b} q) ) p^{q}_{2} &=& -i(\sum_{q^{'}} \ve_{2}^{q^{'}}m_{q^{'}q} - 
\ve_{2}^{q} n_{2} - \ve^{q}_{1}r^{*} ) \nonumber \\
( \pp - i t_{0}(k_{2}v -k_{1}v - 2 \D _{2} +2 \D_{1} ) ) r &=& 
-i\sum_{q} (\ve_{1}^{q}p_{2}^{q*} - \ve_{2}^{q*} p_{1}^{q} ) 
\nonumber \\
\pp n_{j} &=& -i \sum_{q} ( \ve^{q}_{j}p^{q*}_{j} - 
\ve^{q*}_{j}p^{q}_{j})  \nonumber \\
( \pp + it_{0} \O_{b} (q-q^{'} ) ) m_{qq^{'}} & = & -i 
\sum_{j=1,2}( \ve^{q*}_{j}p^{q^{'}}_{j} - \ve^{q^{'}}_{j}p^{q*}_{j}) 
\label{threemb}
\een
where $\ve_{j}^{q}, ~~j= 1,2$ is the amplitude of a double-frequency 
ultrashort pulse and $q = \pm 1$ denote the right(left)-handed 
polarization. Other variables are proportional to the components of 
the density matrix 
\ben
p_{1}^{q} &=& \rho ^{(ba)}_{-q0}\exp [-i(k_{1}x-w_{1}t)]/N_{a},~~
p_{2}^{q} = \rho ^{(bc)}_{-q0}\exp [-i(k_{2}x-w_{2}t)]/N_{a} 
\nonumber \\
n_{1} &=& -\rho _{00}^{(a)}/N_{a}, ~~ 
n_{2} = -\rho _{00}^{(c)}/N_{a} , ~~
m_{qq^{'}} = -\rho^{(b)}_{-q^{'}-q}/N_{a}  \nonumber \\
r &=& - \rho ^{(ca)}_{00}\exp [i (k_{1}-k_{2})x-i(w_{1}-w_{2})t ]/N_{a} 
\een
and $t_{0}$ is a constant with the dimension of time and $N_{a}$ is the 
polulation density of the level $|a >$. 
$2 \D_{1} \equiv w_{1} - w_{ba}, ~ ~ 2 \D_{2} \equiv w_{2} - w_{bc} $ 
measure the amount of detuning from the resonance frequencies. 
The integrability of (\ref{threemb}) comes from its equivalent zero 
curvature formulation with the $4 \times 4$ matrix $U-V$ pair,
\ben
U &=& \pmatrix{-A_{1} - i\l {\bf 1}_{2 \times 2} & -i E \cr 
		  -iE^{\dagger } & -A_{2} + i\l {\bf 1}_{2 \times 2} }
\nonumber \\
&& \nonumber \\
V &=& {i \over 2\l }\pmatrix{ -M & P \cr P^{\dagger } & -N }
\een
where 
\begin{eqnarray*}
E &=& \pmatrix{\ve^{-1}_{1} & \ve^{-1}_{2} \cr \ve^{1}_{1} 
& \ve^{1}_{2} }, ~ 
P = \pmatrix{ p^{-1}_{1} & p^{-1}_{2} \cr p^{1}_{1} & p^{1}_{2} } , ~ 
M = \pmatrix{m_{-1-1} & m_{1-1} \cr m_{-11} & m_{11} } \nonumber \\
&& \nonumber \\
N &=& \pmatrix{n_{1} & r^{*} \cr r & n_{2} }, ~ A_{1} = \pmatrix{a & 0 
\cr 0 & b}, ~ A_{2} = \pmatrix{ x & 0 \cr 0 & y }
\nonumber
\end{eqnarray*}
with
\ben
a &=& {i t_{0} \over 4}(k_{1}v + k_{2}v - 2 \D_{1} - 2 \D_{2} + 4\O_{b} )
\nonumber \\
b &=& {i t_{0} \over 4}(k_{1}v + k_{2}v - 2 \D_{1} - 2 \D_{2} - 4\O_{b} ) 
\nonumber \\
x &=& {i t_{0} \over 4}( -3 k_{1}v + k_{2}v + 6 \D_{1} - 2 \D_{2} ) 
\nonumber \\
y &=& {i t_{0} \over 4}( k_{1}v -3 k_{2}v - 2 \D_{1} +6 \D_{2} )
\label{degthe}
\een
In the context of effective field theory, this corresponds to the case 
where $G/H = SU(4) /S(U(2) \times U(2))$ with the gauge fixing
\be
A = \pmatrix{ A_{1} & 0 \cr 0 & A_{2} }, ~ \Ab = 0
\ee
and
\be
T = -\bar{T} = {i} \pmatrix{ {\bf 1}_{2 \times 2} & 0 \cr 0 & 
-{\bf 1}_{2 \times 2} }.
\ee
Similarly, we may repeat an identification for the case \cite{Bash3}, 
$j_{a} = j_{c} = 1, j_{b} = 0$, and can easily verify that it 
corresponds to the symmetric space $SU(5)/U(4)$. 
Thus, we have shown that various cases of SIT all correspond to the 
effective field theories with specific symmetric spaces, i.e. 
\ben
SU(2)/U(1) & \leftrightarrow & \mbox{nondegenerate two-level} 
\nonumber \\
SU(3)/U(2) & \leftrightarrow & \mbox{degenerate two-level}; 
j_{b} = 0  \rightarrow j_{a} = 1, \nonumber \\ 
&& ~~~~~~~~~~~~ j_{b} = 1 \rightarrow j_{a} = 0, ~
j_{b} = 1 \rightarrow j_{a} = 1 \nonumber \\
(SU(2)/U(1))^{2} & \leftrightarrow & \mbox{degenerate two-level};
j_{b} =1/2 \rightarrow j_{a} = 1/2 \nonumber \\
SU(4)/S(U(2) \times U(2)) & \leftrightarrow & 
\mbox{degenerate three level};
~ j_{a} = j_{c} =  0, ~ j_{b} = 1.
\nonumber \\
SU(5)/U(4) & \leftrightarrow & \mbox{degenerate three level};
~ j_{a} = j_{c} = 1, ~ j_{b} =0. \nonumber \\
&&
\een
Each case corresponds to a special type of symmetric spaces known as 
Hermitian symmetric spaces. Hermitian symmetric space is a symmetric 
space equipped with a complex structure which in our case is given by 
the adjoint action of $T$ up to a scaling. In Sec. 5, the characteristic 
properties of Hermitian symmetric space will be used in obtaining 
infinitely many conserved local integrals. Our association of various 
SIT systems with Hermitian symmetric spaces suggests that to each 
Hermitian symmetric space there may exist a specific SIT with a proper 
adjusting of physical parameters. Especially, the multi-frequency 
generalization of SIT in a configuration of the ``bouquet" type may 
correspond to the symmetric space $SU(n)/U(n-1)$ for an integer $n$. 
However, for large $n$, it requires a fine tuning of physical 
parameters which makes the theory unrealistic.

\subsection{Inhomogeneous broadening}
So far, we have identified various cases of SIT with specific effective 
field theories when inhomogeneous broadening is neglected. In the 
presense of inhomogeneous broadening, due to the microscopic motion 
of atoms, each atom in a resonant medium responds to the macroscopic 
incoming light with different Doppler shifts of transition frequencies. 
Thus, microscopic variables, e.g. the polarization $P$ and the 
population inversion $D$ are characterized by Doppler shifts and they 
couple to the macroscopic variable $E$ through an average over 
the frequency spectrum as given in (\ref{inhomog}). A remarkable 
property of our effective field theory formulation is that it 
incorporates inhomogeneous broadening naturally only with minor 
modifications. The notion of effective potential variable $g$ still 
persists where the microscopic variable $g$ becomes also a function 
of frequency $\xi $, i.e. $g = g(z, \zb , \xi )$. However, the action 
principle in (\ref{action}) is no longer valid despite the use of the 
potential variable $g$. In order to formulate the SIT equation with 
inhomogeneous broadening in terms of the potential variable $g$, we 
relax the constraint equation (\ref{constraint2}) and require only
\be
(\gi \pp g   +\gi A g)_{\bf h} - A  = 0 \ ,
\label{inhocon}
\ee
where the subscript specifies the projection to the subalgebra 
${\bf h}$. The linear equation is given by
\ben
L_{z}\Psi &\equiv & \left( \partial + g^{-1}\partial g +  
g^{-1}Ag - \xi T + \tilde{\lambda }T \right) \Psi = 0 \nonumber \\
L_{\zb }\Psi &\equiv &
\left( \ \bar{\partial} + 
\left< {g^{-1}\bar{T} g \over \tilde{\lambda } - \xi ^{'} 
} \right> \right) \Psi = 0
\label{inholin}
\een
where the constant $\tilde{\lambda }$ is a modified spectral parameter 
which becomes $\lambda + \xi $ in the absense of inhomogeneous 
broadening. The angular brackets denote an average over $\xi^{'}$ as 
in (\ref{inhomog}). As in the previous examples without inhomogeneous 
broadening, we make the same identification of the matrix 
$g^{-1}\partial g +  g^{-1}Ag - \xi T $ with various components of 
macroscopic electric fields which are independent of the microscopic 
quantity $\xi$. This requires the $\xi$-dependence of 
$g(z, \zb , \xi )$ to be such that $g^{-1}\partial g +  g^{-1}Ag - 
\xi T $ is independent of $\xi $. It is easy to see that this 
requirement is indeed satisfied by various integrable SIT systems we 
have considered previously except for the three-level system, where 
we must take $\x = -t_0 \D_1 = -t_0 \D_2$. It means that in order to 
preserve the integrability in the presence of inhomogenuous 
broadening, two detuning parameters of the three-level system must be 
equal.  Note that $\Psi (\tilde{\l} , z, \zb )$ is not a function of 
$\x$. The integrability of the linear equation (\ref{inholin}) becomes
\ben
0 &=& \left[ \partial + g^{-1}\partial g + g^{-1}Ag - \xi T + 
\tilde{\lambda }T \ , \ \bar{\partial} + 
\left< {g^{-1}\bar{T} g \over \tilde{\lambda } - \xi^{'} } \right> 
\right] 
\nonumber \\
&=& - \pb ( g^{-1}\partial g +  g^{-1}Ag - \xi T ) 
+ \left< [T, ~ g^{-1}\bar{T} g ] \right>
\label{inhozero}
\een
where we used the fact that $g^{-1}\partial g + g^{-1}Ag - \xi T $ is 
independent of $\xi $ and also the identity
\be
\pp ( g^{-1}\bar{T} g ) + [ g^{-1}\partial g + g^{-1}Ag , ~ 
g^{-1}\bar{T} g ] = 0.
\label{ident}
\ee
Identifying $g^{-1}\bar{T} g$ with components of the density matrix as 
in the previous cases, we obtain the SIT equation with inhomogenous 
broadening. For example, we may identify $E, P$ and $D$ as in 
(\ref{epdtwo}) so that (\ref{inhozero}) and (\ref{ident}) become the 
SIT equation with inhomogeneous broadening for the nondegenerate 
two-level case given by (\ref{sit}). Note that each frequency $\xi $ 
corresponds to a specific gauge choice of the vector $U(1)$ subgroup. 
Therefore, inhomogenous broadening is equivalent to averaging over 
different gauge fixings of $U(1) \subset H$. This implies that 
inhomogenously broadening can not be treated by a single field 
theory thereby lacking a Lagrangian formulation. It is remarkable 
that nevertheless the group theoretical parametrization of 
various physical variables in terms of an effective potential $g$ 
is still valid. This group theoretical structure underlies the 
integrability of the SIT equation as we will show in Sec. 4.

\subsection{Effective potentials}
One of the main advantage of our effective field theory approach to SIT 
problems is the introduction of a notion of effective potential energy. 
This allows us a systematic understanding of stability of solutions as 
well as the topological nature of soliton solutions.
The effective potential energy term in (\ref{potent}) reveals a rich 
structure of the vacuum of the theory. In general, the potential 
(\ref{potent}) is ``periodic" according to the coset structure $G/H$. 
This results in infinitely many degenerate vacua, which are specified 
by a set of integers, and also soliton solutions - finite energy 
solutions which interpolate between different vacua. In the 
nondegenerate two-level case, the potential term (\ref{potential}) 
becomes a periodic cosine potential (\ref{twopot}) and each degenerate 
vacua are labeled by an integer $n$ as in (\ref{twovac}). Solitons 
interpolating between two different vacua with labels $n_{a}$ and 
$n_{b}$ as $x$ varies from $-\infty $ to $\infty $  are characterized 
by a soliton number $\D n = n_{b} - n_{a}$. 
In order to understand the vacuum structure of the potential for a more 
general $G/H$ case, we first note that the potential term $\mbox{Tr} 
(g T g^{-1} \Tb )$ is invariant under the change $g \rightarrow gh$ 
for $h \in H$. Consequently, we may express the potential term by a 
coset element $m \in G/H$ such as $\mbox{Tr}( m T m^{-1} \Tb ) $, 
where $m$ is given specifically by 
\be
m  = \exp  \pmatrix{ 0 &  B  \cr & \cr  -B^\dagger  &  0} =
\pmatrix{ \cos \sqrt{B B^\dagger} &  
B\sqrt{B^\dagger B}^{-1} \sin \sqrt{B^\dagger B}  \cr & \cr
 -\sin \sqrt{B^\dagger B}  \sqrt{B^\dagger B}^{-1 } B^\dagger  & 
\cos \sqrt{B^\dagger B} }
\label{mmat}
\ee 
The matrix $B$ parametrizes the tangent space of $G/H$. This manifests 
the periodicity of the potential through the cosine and the sine 
functions. In the case of our interest, $B$ is a complex-valued matrix 
of dimension $1 \times 1, ~ 1 \times 2$ and $ 2 \times 2 $ for the coset  
$ SU(2)/U(1), ~ SU(3)/U(2) $ and $ SU(4)/S(U(2) \times U(2)) $ 
respectively. Then, due to the relation: $B \sin \sqrt{B^\dagger B} 
\sqrt{B^\dagger B}^{-1} = \sin \sqrt{B B^\dagger}  
\sqrt{B B^\dagger}^{-1} B $, the potential term reduces to the form
\be
\mbox{Tr} \left(I - 2 \sin^2 \sqrt{B B^\dagger} \right) 
+ \mbox{Tr} \left(I - 2 \sin^2 \sqrt{B^\dagger B} \right) 
\label{pot1}
\ee
for $SU(2)/U(1)$ and $SU(4)/S(U(2) \times U(2))$ cases and
\be
\mbox{Tr} \left(4 I - 6 \sin^2 \sqrt{B B^\dagger} \right) 
+ \mbox{Tr} \left(I - 3 \sin^2 \sqrt{B^\dagger B} \right) 
\label{pot2}
\ee
for $SU(3)/U(2)$ case.
In order to reduce further, we define non-zero eigenvalues of 
${B^\dagger B}$ by  $\f_i ^2,  i=1,.., r \equiv \mbox{rank} 
\{B^\dagger B\}$, which are positive definite and coincide with those 
of ${B B^\dagger}$. With these eigenvalues, the potential term takes 
a particularly simple form, 
\be
a - b \sum _{i} \sin ^2 \f_i ,
\label{repot}
\ee
where the positive constants $a$ and $b$ can be read directly from 
(\ref{pot1}) and (\ref{pot2}). This manifests the periodicity of 
the potential and the infinite degeneracy of the vacuum.  
The minima of the potential (\ref{repot}) occur at $\f_i = 
(n_i + 1/2) \p$  for integer $n_{i}$ thus the degenerate 
vacua are specified by a set integers $(n_{1}, n_{2}, ... ,n_{r})$.
The rank $r$ of ${B^\dagger B}$ is one for $SU(2)/U(1)$ and 
$SU(3)/U(2)$  and two for $SU(4)/S(U(2) \times U(2))$.
Thus solitons for the $SU(4)/S(U(2) \times U(2))$ case, interpolating 
between two vacua $(n_{1a}, n_{2a})$ and $(n_{1b}, n_{2b})$, are 
labeled by two soliton numbers $\D n_{1} = n_{1b} - n_{1a} $ and 
$\D n_{2} = n_{2b} - n_{2a}$. As a specific example, we may write 
the $B$ matrix of $SU(3)/U(2)$ by
\be
B = (-\f \sin \h e^{-i \b} \ \ -\f \cos \h e^{-i \a})
\ee
so that the $m$ matrix (\ref{mmat}) is given by
\be
m = \left( \begin{array}{ccc}
\cos \f & -\sin \f \sin \h e^{-i \b} & -\sin \f \cos \h e^{-i \a} \\
\sin \f \sin \h e^{i \b} & \cos ^2 \h + \cos \f \sin^2 \h & 
-\cos \h \sin \h e^{i\b -i \a} (1 - \cos \f ) \\
\sin \f \cos \h e^{i \a} & -\cos \h \sin \h e^{ i \a - i \b} 
(1- \cos \f ) &
\sin^2 \h + \cos \f \cos^2 \h  \end{array} \right) .
\ee
Then, the potential term becomes 
\be
{\rm Tr}( g T g^{-1} \Tb ) = {\rm Tr}
(m T m^{-1} \Tb )= 6 -9 \sin^2 \f 
\ee
which agrees with (\ref{pot2}).

\section{ Dressing and solitons }
\setcounter{equation}{0}
In order to find exact solutions of the SIT equations with 
inhomogeneous broadening, we first note that the linear equation 
(\ref{inholin}) admits an application of the dressing method. The 
dressing method is a systematic way to obtain nontrivial solutions 
from a trivial one. In our case, we take a trivial solution by 
\be 
g = 1 \mbox{ and }
 \Psi = \Psi^{0} \equiv
 \exp[ - (A -\xi T + \lt T )z -  \left< { \bar{T}  \over \tilde{\l } - 
 \xi^{'} } \right>\bar{z} ] .
\ee
Let $\Gamma $ be a closed contour or a contour extending to infinity 
on the complex plane of the parameter $\lt $ and $G(\lt )$ be a matrix 
function on $\Gamma $.  Consider the Riemann problem of 
$\Psi^0 G(\lt )(\Psi^0)^{-1}$ on $\Gamma $ which consists of the 
factorization 
\be
\Psi^0 G(\lt )(\Psi^0)^{-1} = 
(\Phi_{-})^{-1} \Phi_{+}
\label{factor}
\ee
where the matrix function $\Phi_{+}(z,\zb ,\lt )$ is analytic with $n$ 
simple poles $\m_{1}, ... ,\m_{n}$ inside $\Gamma $ and 
$\Phi_{-}(z,\zb ,\lt )$ analytic with $n$ simple zeros 
$\l_{1}, ... , \l_{n}$ outside $\Gamma$. We assume that none of these 
poles and zeros lie on the contour $\Gamma$ and the factorization is 
analytically continued to the region where $\l \ne \m_{i} ,
\l_{i} \ ; \ i = 1,...,n$.
We normalize $\Phi_{+}, \Phi_{-}$ by  $\Phi_{+}|_{{\tilde \l} = 
\infty }= \Phi_{-}|_{{\tilde \l} = \infty } =1$. 
Differentiating (\ref{factor}) with 
respect to $z$ and $\zb $, one can easily show that
\ben
\pp \Phi_{+} \Phi_+^{-1} -  \Phi_{+} (A-\xi T + \lt T ) 
\Phi_{+}^{-1} &=& \pp \Phi_{-} \Phi_{-}^{-1} - \Phi_{-} (A - \xi T + 
\lt T ) \Phi_{-}^{-1}
\nonumber \\
\pb \Phi_{+} \Phi_+^{-1} - \left< { \Phi_{+} \Tb \Phi_{+}^{-1} \over 
\lt - \xi^{'} }  \right>  &=& \pb \Phi_{-} \Phi_{-}^{-1} - 
\left< { \Phi_{-} \Tb \Phi_{-}^{-1} \over \lt - \xi^{'} } \right>  .
\een
Since $\Phi_{+} (\Phi_{-})$ is analytic inside(outside) $\Gamma$, 
we find that the matrix functions $\bar{U} $ and $ \bar{V} $, defined by
\ben
\bar{U} &\equiv &  -\pp \Phi \Phi^{-1} +  \Phi (A - \xi T + \lt  T )
\Phi ^{-1} - \lt T
 \nonumber \\
\bar{V}  &\equiv &  -( \lt - \xi ) \pb \Phi \Phi^{-1} +  \Phi \Tb 
\Phi^{-1}
\een
where $\Phi = \Phi_{+}$ or $\Phi_{-}$ depending on the region, become
independent of $\lt $. Then, $\Psi \equiv \Phi \Psi^{0}$ satisfies the
linear equation;
\be
(\pp + \bar{U} + \lt T )\Psi = 0 \ , \ (\pb + \left< { \bar{V}  
\over \tilde{\l } - \xi^{'} }  \right> )\Psi = 0 \ .
\ee
Since $\bar{U}, \bar{V}$ are independent of $\lt $, we may fix $\lt $ by 
taking $\lt = \xi $. Define $g$ by $ g \equiv H\Phi^{-1} |_{\lt = \xi } $ 
where $H$ is an arbitrary constant matrix which commutes with 
$T, \bar T$ and $A$. Then, $\bar{U}$ and $ \bar{V}$ become
\ben
\bar{U} &=& g^{-1}\pp g +  g^{-1}Ag - \xi T  \\
\bar{V} &=&  g^{-1}\bar{T}g .
\een
If we further require the constraint condition (\ref{inhocon}) on 
$\Phi^{-1} |_{\lt = \xi } $ such that
\be
 (-\pp \Phi \Phi^{-1} +  \Phi A \Phi ^{-1})_{\bf h} -A =0,
\ee
we obtain a nontrivial solution $g$ and $\Psi $ from a trivial one. 
The nontrivial solution in general describes n-solitons coupled with 
radiation mode. If $G(\lt ) = 1$ in (\ref{factor}), we obtain exact 
n-soliton solutions. This formal procedure may be carried out 
explicitly for each cases of SIT in Sec. 3 and a closed form of 
n-soliton solutions can be obtained. In the following, we give 
explicit expressions of 1-soliton solution of two different cases 
and discuss about their physical properties.
\subsection{Nondegenerate two-level case}
Here, we set $\b =1$ without loss of generality. The 1-soliton solution 
may be obtained either by using 
the above dressing method or by applying the B\"{a}cklund transformation 
directly to the trivial vacuum solution \cite{shin1}. It is given by
\ben
\cos{\vf } &=& {b \over \sqrt{(a-\x )^{2} + b^{2} }} 
\mbox{sech} (2bz -2b C \zb ) \nonumber \\
\h &=& (a-\x )z + (a-\x ) C\zb \nonumber \\
\q &=& - \tan^{-1}[ {a-\x \over b } \mbox{coth} (2bz - 2bC\zb )] - 
2\x z + 2D\zb 
\label{solisol}
\een
where $a, b$ are arbitrary constants and 
\be
C = \left< { 1\over (a-\x^{'} )^{2} + b^{2}} \right> \ ,
 \ D = (a-\x )\left< { 1\over (a-\x^{'} )^{2} + b^{2}} \right> 
- \left< {  a-\x^{'} \over (a-\x^{'} )^{2} + b^{2}} \right> .
\ee
In terms of $E$ as defined in (\ref{csgepd}), 1-soliton is given by
\be 
E = -2ib \  \mbox{sech} (2bz - 2bC \zb ) e^{-2i(az -D\zb  + 
(a-\x )C\zb )} .
\label{esoliton}
\ee
In the sharp line limit of the frequency distribution $f(\xi ^{'}) = 
\d (\xi ^{'} - \xi ) $, 1-soliton retains the same form except 
for the change of constants $C$ and $D$,
\be
C =  { 1\over (a-\x )^{2} + b^{2}}  \ , \ D = 0 .
\ee
The name ``$n$-soliton" is rather ambiguous since even in the sharp 
line limit the number $n$ does not 
necessarily mean the topological soliton number. For example,
when $a=\x$ in the above 1-soliton solution, the solution describes a 
localized pulse configuration which interpolates
between two different vacua of (\ref{twovac}) such that 
\be
\f (x = -\infty) = (n + {1 \over 2}) \p , ~~
 \f (x = \infty) = (n + {1 \over 2}-{b \over |b| }(-1)^{n} ) \p .
\ee
Thus it carries a topological number $\D n = (-1)^{n+1}b/|b| $ and 
becomes a topological soliton. However, when $a \ne \x$, the solution 
interpolates 
between the same vacuum since the peak of the localized solution does 
not reach to the point where $\cos {\vf } = 1$. Therefore, its 
topological number is zero. Nevertheless, it shares many important 
properties, e.g. localization, scattering behavior etc., with the 
topological soliton so as to deserve the name, a ``nontopological 
soliton". One of the important feature of a nontopological soliton is 
that instead of a topological charge, it carries a $U(1)$ charge which 
accounts for stability of a nontopological soliton. The issue of 
$U(1)$ charge and stability will be addressed in the following 
sections. It is remarkable that the nontopological soliton can be 
obtained from the topological one by the local vector transform 
(\ref{gaugetr}). Note that even though the action (\ref{action}) is 
invariant under the gauge transformation, the physical quatities like 
$E, P$ and $D$ are not invariant. The gauge choice (\ref{gfix}) refers 
to the amount of detuning with a detuning parameter $\D \o = \xi $. 
Therefore, by choosing a different gauge condition through the vector 
gauge transformation, we obtain a system with a different amount of 
detuning. In particular, we can obtain a solution $g(\D \o = \x )$ 
of SIT theory with detuning parameter $\xi $ from the solution with 
zero detuning through the transform
\be
g(\D \o = \xi ) = h^{-1} g (\D \o = 0)  h,~~~ h = \exp (\xi z T).
\ee
Or, in case of soliton solution, the topological soliton ( $a-\xi =0$) 
can be mapped to the nontopological one ($ a - \xi \ne 0$).
For this reason, in this paper we will simply call n-solitons for both 
the topological and the nontopological solitons unless otherwise stated 
explicitly.
\subsection{Degenerate three-level case}
We first restrict the degenerate three-level system to the resonant 
case ($\D_{1} = \D_{2} = 0, ~ v=0$) without external magnetic field 
and inhomogeneous broadening. This is equivalent 
to the case where $A=\Ab = 0$ in (\ref{uv}) with identifications 
(\ref{degthe}) in terms of a $4 \times 4$ matrix $g$. 
By applying the B\"{a}cklund transformation for the symmetric space 
sine-Gordon theory in Ref. \cite{bps}, we obtain the 1-soliton 
solution with the parametrization as in (\ref{mmat}),
\be
B = -2 B_{0}\tan ^{-1} \exp ( 2b z + {2 \over b }\zb + const. )
\equiv \f B_{0}
\ee
where $b $ is a constant and $B_{0}$ is a constant $2 \times 2$ 
matrix satisfying 
\be
B_{0}B_{0}^{\dagger }B_{0} = B_{0} .
\ee
If the matrix $B_{0}$ is degenerate, i.e. $\mbox{det}B_{0} = 0$, 
$B_{0}$ can be given in general by
\be
{ i\over \sqrt{1 + |\alpha |^{2}}} \pmatrix{\theta_{1} & \theta_{2} \cr 
\alpha \theta_{1} & \alpha \theta_{2} }
\ee
with complex constants $\alpha , \theta_{1} , \theta_{2}$ satisfying 
$|\theta_{1}  |^{2}  + |\theta_{2}  |^{2}  = 1$. The eigenvalues of 
$B_{0}B_{0}^{\dagger } $ are then zero and one. Therefore, up to a 
global $SU(2)$ similarity  transform of $ B_{0}B_{0}^{\dagger } $, 
this solution corresponds to the (1,0) or (0,1)-soliton as explained 
in Sec. 3.3. This solution has been known as a simulton in earlier 
literatures. Note that the effective potential (\ref{potential}) is 
invariant under the similarity transform. This sounds to be 
contradictory since the similarity transform mixes two different 
emission modes of radiation with different resonance frequencies, 
$w_{ba} \ne w_{bc}$, consequently different transition energies. 
However, in considering the degenerate three-level case, we have 
assumed that the oscillator strengths are equal,
\be
k_{1}|d_{ba}|^{2} = k_{2}|d_{bc}|^{2},
\ee
for incident wave numbers $k_{1}, k_{2}$ and the reduced dipole momenta 
$d_{ba} , d_{bc}$  \cite{Bash2}. This assumption, together with a 
mediatory role of off-diagonal components of the density matrix, the 
similarity transform in fact becomes an invariance of the potential.

For the nondegenerate $B_{0}$, we can take $B_{0}$ as an arbitrary 
$U(2)$ matrix so that $B_{0}B_{0}^{\dagger } = {\bf 1}_{2\times 2}$ 
and the corresponding solution is the (1,1)-soliton in Sec. 3.3.  
This is different from (1,0) or (0,1)-soliton and can not be reached by 
the similarity transform since the similarity transform preserves
eigenvalues of $B_{0}B_{0}^{\dagger }$. In fact, they are topologically 
distinct and separated by an infinite potential energy barrier.
Finally, physical quantities can be obtained from $g$ through the 
identification (\ref{uv}). Explicitly,  we find $E$, $P$ and $M$ in 
(\ref{degthe}) to be
\ben
E &=& i B_{0} \pp \f = -2 i \eta B_{0} \mbox{sech}(2 \eta z + 
{2 \over \eta} {\bar z} + const. ) \nonumber \\
P &=& -2 B_{0} \sin 2 \f, \nonumber \\
 M &=& -N = -2 {\bf 1}_{2 \times 2} \cos 2 \f
\een
respectively. Inclusion of detuning and external magnetic effects can 
be done easily by a gauge transform;
\be
E \rightarrow H_{1}^{-1}EH_{2}, ~ M \rightarrow H_{1}^{-1}MH_{1}, ~ 
P \rightarrow H_{1}^{-1}PH_{2}, ~ N \rightarrow H_{2}^{-1}NH_{2}
\ee
where $H_{1}, H_{2}$ are given by $A_{1}=H_{1}^{-1}\pp H_{1} , 
~ A_{2}=H_{2}^{-1}\pp H_{2}$ for $A_{1}, A_{2}$ in (\ref{degthe}).

\section{Symmetries}
\setcounter{equation}{0}
The effective field theory action (\ref{action}) reveals various types 
of symmetries of SIT.
As an integrable field theory, at least classically, it possesses 
infinitely many conserved local integrals. These conservation laws can 
be extended to the inhomogeneously broadened case without difficulty. 
Some explicit local integrals for the nondegenerate two-level SIT have 
been given in earlier literatures \cite{lamb2}. Here, we present a 
systematic way to find local integrals for the general $G/H$ case 
and present a few explicit examples. Another important symmetry of 
(\ref{action}) is the global $U(1)$ axial symmetry; $g \rightarrow hgh$ 
for a constant element $h \in U(1) \subset H$. 
This type of symmetry has been previously unknown and is one of the 
outcomes of our effective field theory approach. We show that the $U(1)$ 
symmetry and its corresponding charge plays a crucial role in pulse 
stability problem.

Besides the aforementioned continuous symmetries, the action 
(\ref{action}) also possesses two distinct types of discrete 
symmetries; the chiral and the dual symmetries. 
Discrete symmetries relate two different solutions of the SIT equation. 
In particular, the dual symmetry relates the ``bright" soliton with the 
``dark" soliton of SIT.  These discrete symmetries reflect the ubiquitous 
nature of the action (\ref{action}) without the potential term, as an 
action for the coset conformal field theory and also a bosonized version 
of 1+1-dimensional free fermion field theories.

\subsection{Conserved local integrals}
In the previous section, we have shown that the linear equation with a 
spectral parameter yields exact soliton solutions through the dressing 
procedure. The same linear equation can be employed to construct 
infinitely many conserved local integrals. In this section, using the 
linear equation as well as the properties of Hermitian symmetric spaces, 
we present a systematic way of finding such integrals  for various cases 
of SIT introduced earlier.  

We first recall some relevant mathematical facts on Hermitian symmetric 
space \cite{helgason}. A symmetric space $G/H$ is a coset space with the 
Lie algebra commutation relations among generators of associated Lie 
algebras such that
\be
[ \bf{h} \ , \ \bf{h} ] \subset \bf{h} \ ,\ [ \bf{h} \ , \ \bf{m} ] 
\subset \bf{m} \ , \ 
[\bf{m} \ , \ \bf{m} ] \subset \bf{h} \ ,
\ee
where $\bf{g}$  and $\bf{h}$ are Lie algebras of $G$ and $H$ and 
$\bf{m}$ is the vector space complement of $\bf{h}$ in $\bf{g}$, i.e. 
\be
\bf{g} = \bf{h} \oplus \bf{m}.
\label{decom}
\ee
Hermitian symmetric space is a symmetric space equipped with a 
complex structure. In general, such a complex structure 
is given by the adjoint action of $T_{0}$ on ${\bf m}$ up to a scaling, 
where $T_{0}$ is an element belonging to the Cartan subalgebra of 
${\bf g}$ whose stability subgroup is $H$. In our case, $T_{0}$ is 
precisely the $T$-matrix given in Sec. 3. Namely, with a suitable 
normalization of $T$, we have
\be
T \in {\bf h} \ \  , \ \ [T\ , \ {\bf h}]=0 \ \ \mbox{ and } 
\ \ [T\ ,\ [T\ ,\ a]]= - a \ \mbox{ for any } a \in {\bf m}.
\ee
We decompose an algebra element $\j \in {\bf g} $ according to 
(\ref{decom}),
\be
\j = \j_{h} + \j_{m} .
\ee
Such a decomposition could be extended to a group element  $\J 
\in G = SU(n)$ if we substitute the commutator by a direct matrix 
multiplication and add an identity element $h_{0} = I $ to the 
subalgebra ${\bf h}$, i.e.
\be
\J = \J_{h} + \J_{m} , ~~~ \J_{h} \J_{h} \subset \J_{h} , ~~ 
\J_{h}\J_{m} \subset \J_{m} , ~~ \J_{m} \J_{m} \subset \J_{h} .
\ee
In other words, any unitary $n \times n$ matrix can be expressed 
as a linear combination of $SU(n)$ generators and the identity element 
$h_0$ such that 
\be
\J = \J_{h} + \J_{m} = \sum_{a=0}^{\mbox{dim } \bf h} \J ^{a} h_{a} + 
\sum_{b=1}^{\mbox{dim } \bf m} \J ^{b} m_{b} .
\ee
In order to solve the linear equation (\ref{inholin}) recursively, we 
expand $\J$ in terms of a power series in $\tilde {\l}$,
\be
\J \exp (-{\tilde {\l}} T z) =\sum _{i=0} ^{\infty} {1 \over {\tilde 
{ \l}}^{i} } \Phi_{i}, ~ \mbox{ where } ~ 
\F_i = \sum_{a=0}^{\mbox{dim } \bf h} \F_{i} ^{a} h_{a} + 
\sum_{b=1}^{\mbox{dim } \bf m} \F_{i}^{b} m_{b} \equiv \cC _{i} + 
\cD _{i} .
\ee 
We also introduce the abbreviation:
\ben
\E &\equiv&  {\bar U} =g^{-1} \pp g +  g^{-1} A g 
- \x T \subset  \bf{m} \nonumber \\
\left< {\bar V} \right>_l &\equiv&
\left< g^{-1} {\bar T} g \right>_l = 
\left< g^{-1} {\bar T} g (-\x^{'})^l \right> = D_l +P_l, \ \ 
D_l \subset {\bf h},
\ \ P_l \subset \bf{m},
\een
so that the linear equations become
\be
(\pp + \E ) \F_{i} - [T\ ,\ \F_{i+1}]=0
\label{itcon1}
\ee
and 
\be
\pb \F_{i} + \sum_{l=0}^{i-1} (D_{i-l-1} + P_{i-l-1}) \F_l =0.
\label{itcon2}
\ee
Then the $\bf{m}$-component of (\ref{itcon1}) is
\be
\pp \cD _{i-1} + \E \cC _{i-1} -[T\ ,\ \cD _{i}]=0 ,
\ee
which can be solved for $ \cD _{i}$ by applying the adjoint action 
of $T$, 
\be
\cD _{i} = - [T\ ,\ \pp \cD _{i-1}]-[T\ ,\ \E ]\cC _{i-1}.
\ee
$\cC _{i}$ can be solved similarly from the $\bf{h}$-component 
of (\ref{itcon1}) and (\ref{itcon2}) such that
\be
\cC _{i} = - \int \E \cD _{i} dz 
- \sum_{l=0}^{i-1} \int (D_{i-l-1} \cC _l + P_{i-l-1} \cD _l ) d \bar z.
\ee
These recursive solutions of $ \cC _{i} $ and $ \cD _{i} $ can be 
determined completely with appropriate initial conditions. For example, 
if we choose an initial condition which is consistent with the recursion 
relation for $i \le 0$,
\be
\cC _0 = I \ \ \ , \ \ \ \cD _0 =0,
\ee
we find for the first few explicit cases in the series,
\be
\cC _1 = \int \E[T\ ,\ \E ] dz -\int D_0 d \bar z , ~~~
\cD _1 = -[T\ ,\ \E ] 
\ee
and 
\ben
\cC _2 &=& \int (\E \pp \E + \E [T\ ,\ \E ] \cC _1 ) dz
+ \int (-D_1-D_0 \cC _1 + P_0[T\ ,\ \E ]) d \bar z 
\nonumber \\ 
\cD _2 &=& - \pp \E - [T\ ,\ \E ] \cC _1  .
\een
Finally, the consistency condition: $\pp \pb \cC _{i} = \pb \pp 
\cC _{i}$ gives rise to infinitely many conserved local currents,
\be
J_{i} \equiv \pp \cC _{i} = - \E \cD _{i} 
, ~~~~ {\bar J}_{i} \equiv \pb \cC _{i} =  
- \sum_{l=0}^{i-1}  (D_{i-l-1} 
\cC _l + P_{i-l-1} \cD _l ) ,
\ee
satisfying $\pp {\bar J}_{i} =\pb J_{i}$. A few examples are
\be
J_1 = \E [T\ ,\ \E] \ \ , \ \ {\bar J}_1 = -D_0
\ee
\be
J_2 = \E \pp \E + \E [T\ ,\ \E ] \cC _1 \ \ , \ \ 
{\bar J}_2 = -D_1-D_0 \cC _1 + P_0[T\ ,\ \E ].
\ee
The first current $J_{1} , ~{\bar J}_{1}$ gives rise to the energy 
conservation law.
With the repetitive use of the properties of the Hermitian symmetric 
space, it can be easily checked that these conservation laws are indeed 
consistent with the equations of motion (\ref{inhozero})(\ref{ident}), 
which in the present convention take a particularly simple form:
\ben
\pb \E - [T ~ , ~ P_0] &=& 0 \nonumber \\
\pp D_{i} + [\E ~,~ P_{i}] &=& 0 \nonumber \\
\pp P_{i} + [\E ~, ~ D_{i} ]-[T\ ,\ P_{i+1}] &=& 0.
\een
In general, the conserved current contains nonlocal terms. 
These nonlocal terms may be dropped out by taking the $T$-component of 
the currents. For instance, the $T$-component of the ``spin-2" current 
conservation is
\be
\pb {\rm Tr} (T \E \pp \E ) = \pp {\rm Tr} (T P_0[T\ ,\ \E ]-T D_1)
\ee
which obviously does not contain nonlocal terms. 
In order to demonstrate the above procedure more explicitly, we take 
the $SU(2)/U(1)$ case as an example. 
In this case, relevant $2 \times 2 $ matrices are given by 
\be                                                              
\E = \pmatrix{ 0 & -E \cr E^{*} & 0}, ~~~ T = {1 \over 2}\pmatrix{ i &
0 \cr 0 & -i } 
\ee
and we introduce the notation
\ben
 \left< g^{-1} \Tb g \right> _{l } &=& \left< g^{-1} \Tb g (-\x^{'} )^{l} 
 \right> 
= -i \pmatrix{ \left< D(\xi^{'} ) (-\xi^{'} )^{l} \right> & 
\left< P(\xi^{'} ) (-\xi^{'} )^{l} \right> \cr \left< P^{*}(\xi^{'} ) 
(-\xi^{'} )^{l} \right>  &
  - \left< D(\xi^{'} ) (-\xi^{'} )^{l} \right> } \nonumber \\
&\equiv &  -i \pmatrix{ D_{l} & P_{l} \cr P_{l}^{*} & -D_{l} }
\een
Let
\be
\Psi \exp(-\lt T z ) = \sum_{i=0}^{\infty } {\Phi_{i} \over \lt ^{i}} 
\ ;  \ \Phi_{i} \equiv \pmatrix{ p_{i} & q_{i} \cr r_{i} & s_{i} }
\ee
so that the linear equation changes into
\be
(\pp + \pmatrix{ 0 & -E \cr E^{*} & 0 } ) \Phi_{i} - [T \ , \ 
\Phi_{i+1} ] = 0
\ee
and 
\be
\pb \Phi_{i} + \sum^{i-1}_{l=0} \left< g^{-1} \Tb g \right> _{i-l-1 }
\Phi_{l} = 0.
\ee
These equations can be solved iteratively in component,
\ben
q_{i} &=& {1 \over 2i} (\pp q_{i-1} - Es_{i-1} ) \\
r_{i} &=& -{1 \over 2 i} (\pp r_{i-1} + E^{*} p_{i-1} ) \\
p_{i} &=& \int Er_{i} dz + i\sum_{l=0}^{i-1}\int (D_{i-l-1}p_{l} + 
P_{i-l-1}r_{l} )d\bar{z} \\
s_{i} &=& -\int E^{*} q_{i} dz + i\sum_{l=0 }^{i-1}\int ( -D_{i-l-1} 
s_{l} +P^{*}_{i-l-1}q_{l} )d\bar{z}
\een
together with the initial conditions:
\be
p_{0} = s_{0} = -2i , \ r_{0} = q_{0} = 0.
\ee
The consistency: $\pp \pb p_{m} - \pb \pp p_{m} = 0$ leads to the 
infinite current conservation laws, $\pp \bar{J}_{i} + \pb J_{i} =0$, 
or 
 \be
i\pp \sum_{l=0}^{i-1} (D_{i-l-1}p_{l} + 
P_{i-l-1}r_{l} ) - \pb ( Er_{i} ) = 0 .
\label{current}
\ee
The consistency condition, $\pp \pb s_{i} - \pb \pp s_{i} = 0$, gives 
rise to the complex conjugate pair of (\ref{current}). 
A few explicit examples of conserved currents are
\ben
\bar{J}_{1} &=& -2D_{0} \nonumber \\ 
J_{1} &=& EE^{*} \\
\bar{J}_{2} &=& 4iD_{1} - 2P_{0}E^{*}  \nonumber \\ 
J_{2} &=& E\pp E^{*} \\
\bar{J}_{3} &=&  -2P_{0}\pp E^{*} + 8D_{2} + 4i E^{*} P_{1} \nonumber \\ 
J_{3} &=& E\pp ^{2} E^{*} + (EE^{*})^{2} \\
\bar{J}_{4} &=& -16iD_{3} + 8E^{*} P_{2} + 4iP_{1}\pp E^{*} -2P_{0} 
\pp^{2}E^{*} - 2P_{0}E^{*}|E|^{2} \nonumber \\ 
J_{4} &=& E\pp ^{3} E^{*} + |E|^{2} \pp |E|^{2} + 2E |E|^{2} \pp E^{*}
\een
\subsection{$U(1)$ symmetry}
One of the properties of the Wess-Zumino-Witten action (\ref{wzw}) is 
the global axial vector gauge symmetry, i.e. 
\be
S_{WZW}(fgf) = S_{WZW}(g)
\ee
for a constant $f \in G$.
The extra terms in the deformed action (\ref{action}) break this global 
axial symmetry in general. Nevertheless, there remains at least an 
unbroken $U(1)$-axial symmetry given by $g \rightarrow hgh$ for 
$h= \exp(\g T) \in U(1)$. This $U(1)$-invariance results in a conserved 
charge according to the Noether method. Even though a general expression 
for the conserved charge should be possible, in practice it requires an 
explicit parametrization of the group variable $g$. Thus, for the sake 
of brevity, here we will restrict ourselves only to the $SU(2)/U(1)$ 
case. In this case, the global axial transform is given by
\begin{equation}
\eta \rightarrow \eta + \g \ \   \mbox{ for } 
\g  \  \mbox{constant} .
\end{equation}
The corresponding Noether currents and the associated axial charge are 
\begin{eqnarray} 
J &=& {\cos^{2}{\varphi } \over \sin^{2}{\varphi }}\partial \eta  \ \ , 
\ \ 
\bar{J} = {\cos^{2}{\varphi } \over \sin^{2}{\varphi }}\bar{\partial} 
\eta \nonumber \\
Q &=& \int_{-\infty }^{\infty }dx (J + \bar{J}) 
\label{charcur}
\end{eqnarray} 
where, owing to the complex sine-Gordon equation 
(\ref{cs})-(\ref{constraint}), $J$ and $ \bar{J}$ can be shown to 
satisfy the conservation law: $  \partial \bar{J} + \bar{\partial} 
J = 0$. In particular, the axial charge of the nontopological 1-soliton 
in (\ref{solisol}) is
\begin{equation}
Q_{\mbox{1-sol}} = c \tan^{-1}{|b| \over a- \x} .
\end{equation}
The charge of the topological soliton is not 
well defined \cite{shin1}.
Stability of nontopological solitons can be proved either by using 
conservation laws in terms of charge and energy as given in  
\cite{shin1}, or by studying the behavior against small fluctuations 
which will be explained in Sec. 6 \cite{shin3}. 

\subsection{Discrete symmetries}
Besides continuous symmetries, the action (\ref{action}) also reveals 
discrete symmetries of SIT, {\it the chiral symmetry} and 
{\it the dual symmetry}. They are manifested most easily in the gauge 
where $A=\Ab =0$. Extensions to different gauges, e.g. the off-resonant 
case which requires a different gauge fixing as in (\ref{gfix}), 
can be made by the vector gauge transform in (\ref{gaugetr}). 

One peculiar property of the action (\ref{action}) is its asymmetry 
under the change of parity $z \leftrightarrow \zb $. This is because 
the Wess-Zumino-Witten action (\ref{wzw}) is a sum of the parity even 
kinetic term and the parity odd Wess-Zumino term thereby breaking parity 
invariance. In the SIT context, broken parity is due to the slowly 
varying enveloping approximation which breaks the apparent parity 
invariance of the Maxwell-Bloch equation. Nevertheless, the action 
(\ref{action}) is invariant under the chiral transform
\be
z \leftrightarrow \zb \  
, \ g \leftrightarrow g^{-1} ~ (\mbox{ or } \h \leftrightarrow -\h 
\ , \ \vf \leftrightarrow -\vf )
\label{chiral}
\ee
which may be compared with the $CP$ invariance in the context of 
particle physics. Thus, parity invariance is in fact not lost but 
appears in a different guise, namely the chiral invariance. This 
chiral symmetry relates two distinct solutions, or itgenerates a new 
solution from a known one. For example, under the chiral transform 
(\ref{chiral}), the 1-soliton solution (\ref{solisol}) in the resonant 
case $(\xi = 0)$ becomes again a soliton but with the replacement of 
constants $a, b$ by  
\be
a \rightarrow -{a \over a^{2} + b^{2}} \ , \ 
b \rightarrow { b \over a^{2} + b^{2}}.
\ee
This implies the change of pulse shape and the change of pulse velocity 
by $v \rightarrow c - v$. The current and the charge also change into
\be
J \rightarrow -\bar{J} \ , \ \bar{J} \rightarrow - J \ , \ Q 
\rightarrow -Q .
\ee
It is remarkable that the velocity changes from $v$ to $c-v$ unlike the 
usual parity change $v \rightarrow -v$.

The other type of discrete symmetry of the action (\ref{action}) is
the dual symmetry of the Krammers-Wannier type: \cite{shin1}
\be
\b \leftrightarrow -\b \ , \ 
g \leftrightarrow  i\s g  
\ee
where $\s$ is a constant matrix with a property $ \s T + T\s   =0$.
For example, $\s = \s_1$ of Pauli matrices for the $SU(2)/U(1)$ case.
This rather unconventional symmetry, also the name, stems from the 
ubiquitous nature of the action (\ref{action}), i.e. it also arises as a 
large level limit of parafermions in statistical physics and the above 
transform is an interchange between the spin and the dual spin 
variables \cite{park}. In general, the change of the sign of $\b $ 
makes the potential upside down so that the degenerate vacua becomes 
maxima of the potential and vice versa. Therefore, the dual transformed 
solutions are no longer stable solutions. This allows us to find a 
localized solution which approches to the maximum of the potential 
asymptotically (so called a ``dark" soliton). In practice, the dark 
soliton for positive $\b $ can be obtained by replacing $\b 
\rightarrow -\b \ , \ \zb \rightarrow -\zb $ in the ``bright" soliton 
(usual one) of the negative $\b $ case. For example, we obtain the 
dark 1-soliton for the $SU(2)/U(1)$ case as follows:
\ben
\cos{\vf }e^{2i\h } &=& - {b \over \sqrt{(a-\x)^{2} + b^{2}}}
\mbox{tanh} (2bz + 2bC\zb ) - i{a-\xi \over \sqrt{(a-\xi )^{2} 
+ b^{2}}} \nonumber \\       
\q &=& -2(a- \x) (z- C\zb ) -2\xi z .
\label{darksol}
\een

\section{Stability}
\setcounter{equation}{0}
The physical relevance of a soliton number is that it accounts for 
the stability of solitons against ``topological" (soliton number 
changing) fluctuations. Note that any finite energy solution must 
approach to one of the degenerate vacua at both ends of the $x$-axis 
thus carries a specific soliton number. In general, solutions are made 
of soliton, antisoliton, breather and radiation modes where each modes 
carry soliton number 1, -1, 0 and 0 respectively. The overall soliton 
number is a conserved topological charge which
satisfies the so-called superselection rule, i.e. it cannot be 
changed during any physical processes due to the infinite potential 
energy barrier between any two finite energy solutions with
different topological charges. This infinite energy barrier results 
from the infinite length of the $x$-axis despite the finite potential 
energy density per unit length. In the case of sine-Gordon solitons, 
the ``area" of the McCall and Hahn's theorem just enumerates 
the soliton number, i.e. 
\ben
\int_{-\infty }^{\infty } 2E dt &=& \int_{-\infty }^{\infty } 2\pp 
\vf dt = 2\vf(t=\infty ) -2\vf(t=-\infty ) \nonumber \\
&=& 2\vf (x = - \infty ) - 2\vf (x= \infty ) ( = -2\p \D n )
\een
where the equality in the last line holds for localized solutions. 
However, this coincidence is only for a specific case where $E$ is 
real and inhomogeneous broadening is ignored. As was shown in Sec. 2, 
the inclusion of a phase degree of freedom into $E$ makes $E$ complex 
so that the time area of $2E$ becomes complex too, thus it loses its 
meaning as a  soliton number. 

Inhomogeneous broadening also spoils the notion of a soliton number. 
In Sec. 3, we have noted that the effective potential variable $g$ 
becomes a microscopic variable depending on the frequency $\xi $ and 
atomic variables $P$ and $D$ also depend on $\xi $. This implies 
that the soliton number is also a $\xi$-dependent microscopic 
variable. However, the macroscopic field $E$, also given in terms 
of $g$, is independent of $\xi $. In other words, the microscopic 
variable $g$ arranges the $\xi $-dependence in such a way that the 
resulting $E$ field becomes independent of $\xi $. This is exemplified 
by the 1-soliton solution in (\ref{solisol})-(\ref{esoliton}).
Thus, inhomogeneous broadening in general requires $E$ to be a function 
of ``frequency $\xi $ averaged" coefficients so that it does not carry a 
topological soliton number. In this regard, the McCall and Hahn's area 
theorem is different from the topological stability argument and 
$2n\pi $ pulses should be distinguished from $n$-solitons. It is 
remarkable that the area theorem provides a stability argument even 
in the absence of the topological argument. In fact, the proof of the 
area theorem relied crucially on the averaging over the frequency 
$\xi$ of detuning associated with inhomogeneous broadening.   
However, one serious drawback of the area theorem is that it applies 
only to a very restricted case, i.e. the case of real $E$ (neglecting 
frequency modulation effect) and the symmetric frequency distribution. 
Presently, a more general area theorem including frequency modulation 
is not known and finding such a theorem would be one of the 
most important problem in the theory of self-induced transparency. 

For the rest of the section, we consider a restricted case which 
however generalizes the area theorem to a certain extent. We consider 
the $SU(2)/U(1)$ case without inhomogeneous broadening and prove the 
area stability around the 1-soliton solution (topological or 
nontopological) using a perturbative argument similar to the one given 
in \cite{lamb2}. This shows that the reshaping of a pulse also  occurs 
in the off resonant case ($a-\xi \ne 0$) but slower than in the resonant 
case. We argue that the $U(1)$ charge of optical pulses introduced 
in earlier sections could be used also in generalizing the area theorem. 
In particular, we prove the $U(1)$ charge stability using the local 
conservation law. These results are in good agreement with numerical 
results obtained earlier \cite{diels}. 

Consider the bright soliton (\ref{solisol}) which describes a coherent 
pulse propagating in the attenuator. In order to overcome the difficulty 
with the complexity of the usual area function for complex $E$, we 
regard $\vf $ of the complex sine-Gordon equation  as a ``modified" 
area function. Without loss of generality, we may take the asymptotic 
time behavior of $\vf $ by
\ben
\vf(t= -\infty,x) =-\p /2 , ~~ \vf(t= \infty,x) =\p /2 ~ \mbox{ for } 
~ a= \x \nonumber \\
\vf(t = -\infty,x) = \vf(t = \infty,x) = -\p /2 ,~ \mbox{ for }~ a \ne 
\x ,
\een
so that the modified area 
\be
A= \int^{\infty }_{-\infty } 2 \pp \vf dt 
\ee  
of the topological soliton ($a-\x =0$) is $2\pi $ while that of the 
nontopological soliton is zero. Now assume that the system was initially 
in the vacuum state ($\vf (t= -\infty,x) =-\p /2$). If the solution is 
perturbed around a soliton so that near the trailing edge of the pulse 
($t >> 1 $), the modified area function changes by 
$\d \vf = \e $ for small $\e $, i.e. $\vf (t>>1) = \pm \pi /2 + \e$. 
Then, the complex sine-Gordon equation for $t >>1$ becomes
\be
\pb \pp \vf + 4 {b^2 \over (a-\x)^2+b^2} \e = 0
\label{recovery}
\ee
where we have neglected the variation of $\eta $ which contributes only  
to the order $\e^{2}$.  
This shows that if the modified area is greater than $2\pi $ 
(or zero) by the amount $\e > 0$, then $\pb \pp \vf < 0$ so that the 
field $\pp \vf $ at the trailing edge tends to decrease along the  
$\zb = x/c$ axis to recover a total modified area $2\pi $ (or zero). 
On the other hand, if the perturbation is such that $\e <0 $, then 
$\pb \pp \vf > 0$ and the field at the trailing edge increases. 
Therefore, the total modified area tends to remain $2\pi $ or zero. 
Moreover, (\ref{recovery}) shows that the recovery of area is faster 
in the resonant case ($a= \x$ ) than in the off-resonant 
case ($a \ne \x $) which agrees precisely with the 
numerical result \cite{diels}. 
In fact, the recovery of the modified area is acompanied by a 
stronger pulse reshaping recovery to that of a soliton. Here, 
we only point out that the stability of a soliton shape against 
fluctuations which preserve the modified area may be shown by making 
a modification of the Lamb's proof using the Liapunov 
function \cite{lamb2}, and also by proving the stability of higher 
order conserved charges as discussed below. In case of the dark 
soliton (\ref{darksol}) which describes a coherent pulse in the 
amplifier, the system is initially in the upper level 
($\vf (t= -\infty , x) = 0$). In such a case, one can easily see that 
the above consideration of perturbation results in the instability of 
a soliton configuration.

It was suggested previously that pulse reshaping arises due 
to the stability of all higher order conserved integrals against 
small fluctuations and each integrals behaves like an 
``area" \cite{AKN}.  
In order to explain the stability of conserved integrals, 
we apply the method introduced in \cite{AKN} to the 
$U(1)$-charge conservation law:
\be
{\pp \over \pp t} [\cot^2 \f ( \pp + \pb ) \h ] + c { \pp \over \pp x}
[\cot^2 \f \pp \h ] = 0.
\label{chargecon}
\ee
Introduce the $U(1)$ charge in terms of a ``time area",
\be
A(x) \equiv \int _{- \infty } ^{+ \infty } [c \cot ^2 \vf \pp \h ] dt,
\ee
so that (\ref{chargecon}) becomes
\be
{dA \over dx} = - \cot^2 \vf (\pp + \pb) \h |_{t= -\infty}^{t= + \infty}.
\ee
The boundary contribution from the 1-soliton is zero thus 
the $U(1)$-charge $A(x)$ is conserved in space, i.e. $dA /dx = 0$.
The physical meaning of the $U(1)$ charge is clear; since 
$\pp \h = a - \x = a+ w_{o} -w$ for the 1-soliton which expresses 
frequency detuning and $\cot ^2 \vf $ is peaked around the soliton, 
the $U(1)$-charge measures precisely the amount of detuning. 
We caution that, as already noted by Lamb \cite{lamb}, the carrier 
frequency $w_{o}$ shifts to $w_{o} + a $ in the presense of a soliton 
so that the detuning parameter is given by $a+ w_{o} -w$. 
Now, if the solution is perturbed around the soliton such that near 
the trailing edge of the pulse, 
\ben
\f (t >> 1 , x)
&=& \pm \p /2 + \e (x) \nonumber \\
\h (t >> 1 , x) &=& (a-\x )(t-{x \over c} - 
{ 1 \over (a-\x )^2 + b^2 } {x \over c}) + \d (x) 
\een 
for small parametric functions $\e (x) $ and $\d (x)$. To the leading 
order, the variation of the $U(1)$-charge then becomes
\be
{d\d A \over dx } = -(a-\x ) \left( 1 + { 1 \over (a-\x )^2 + b^2 } 
\right) \e ^2 .
\ee
This shows that the detuning by a higher frequency, i.e. $a-\x >0 $ 
reduces $A$ for increasing $x$ while the lower frequency 
detuning does exactly the opposite. Since the conserved charge $A$ of 
the 1-soliton is $c \tan^{-1} [ |b|/(a-\x)]$, it can be concluded 
that the absolute value of $A(x)$ decreases monotonically along the 
$x$-axis and converges to a constant charge of the soliton. Note that 
the recovery of $|A|$ value to that of the soliton is slower than the 
area case since it is of the order $\e^2$. The decreasing behavior of 
$|A|$ is in good agreement with the numerical work \cite{diels} which 
showed that the frequency of the optical pulse is pulled towards the 
transition frequency and reaches to a constant value along the $x$-axis.
Thus, the $U(1)$-charge stability provides a generalization of the area 
theorem in the presense of frequency detuning. This type of stability 
argument might be extended to the case of other conserved integrals. 

In the presence of inhomogeneous broadening, the $U(1)$ conservation 
breaks down unlike the local  conservation laws in Sec. 5.1. It 
introduces an anomaly term $M$ in the $U(1)$-current conservation 
such that $\partial \bar{J} + \bar{\partial} J = M$ for $J, \bar{J}$ 
in (\ref{charcur}) and
\begin{eqnarray} 
M &=& 2\cot{\varphi } [ \ \cos (\theta - 2\eta ) 
< \sin (\theta - 2\eta ) \sin {2\varphi } > \nonumber \\ 
&-& 
\sin (\theta - 2\eta ) < \cos (\theta - 2\eta ) \sin {2\varphi } >  
-(\cot^{2}{\varphi }\bar{\partial} \eta  + {1\over 2}\bar{\partial} 
\theta )\partial \varphi   \ ] .
\end{eqnarray} 
This anomaly vanishes in the sharp line limit due to the constraint 
(\ref{constraint} ). It also vanishes in the case of 1-soliton and the 
charge remains conserved. This behavior may be compared with the 
conserved area of topological solitons in the presence of inhomogenous 
broadening. The area theorem of McCall and Hahn proves that inhomogeneous 
broadening changes the pulse area until it reaches to those of 
$2n \pi $ pulses. It remains as an important open question whether 
one could prove a generalized area theorem of pulse stability including 
frequency modulation by making use of $U(1)$ charge and anomaly.

\section{Discussion}
\setcounter{equation}{0}
In this paper, we have presented a field theoretic formulation of SIT. 
Various cases of SIT have been associated with specific symmetric 
spaces $G/H$ and a systematic group theory approaches to the SIT problem 
has been made. In doing so, the introduction of a matrix potential 
variable $g$ was an essential step. One immediate question is about 
the generality of such a group variable in the description of nonlinear 
optics problems. In the nondegenerate two-level case, the $SU(2)$ 
variable $g$ was introduced by solving the constraint which expresses the 
conservation of probability. In more general multi-level cases, the 
complexity of atomic states requires bigger groups $G$ than $SU(2)$. 
However, not all degrees of freedom associated with bigger groups 
become physical degrees of freedom, they are constrained by the Bloch 
equation which requires a restriction of $g$ e.g. as in (\ref{epdtwo}). 
That is, the microscopic atomic variables couple to the macroscopic 
electric field selectively through dipole transitions which suppresses 
the excitation of certain degrees of freedom. This reduction of 
variables has been accomplished by considering the $G/H$ coset structure. 
It should be emphasized that the effective field theory we introduced 
is not an ordinary sigma model where $g$ takes value in the coset $G/H$. 
In the present case, a coset structure is imposed only through a 
nontrivial Lagrange multiplier.\footnote{ More correctly, it 
corresponds to the coset $LG/LH$ of loop groups $LG$ and $LH$ rather 
than $G/H$, see Ref. \cite{qpark} } Unfortunately, this coset theoretic 
description is not valid for all cases of SIT. For example, the 
particular way electric field interacts with two-level degenerate states 
in the $2 \rightarrow 2 $ transition does not allow a nice coset 
structure, but only described by the double sine-Gordon equation when 
further restrictions are made. Thus, a systematic application of group 
theory to the general SIT problems still remains as an interesting open 
question. 

On the other hand, the group theoretical approach is not restricted to 
the SIT problem only. The nonlinear Schr\"{o}dinger equation, which is 
the governing equation for optical soliton communication systems, can 
be also generalized according to each Hermitian symmetric 
spaces \cite{fordy}. In fact, both SIT and the nonlinear Schr\"{o}dinger 
equation share the same Hamiltonian structure and can be combined 
together. We will consider this case and its physical applications 
in a separate paper.
 
Finally, our approach provides a vantage point to the quantum SIT 
problem as well as the quantum optics itself. A direct quantization 
of SIT using the quantum inverse scattering has been made by Rupasov 
and a localized multiparticle state has been found and compared 
with a quantum soliton \cite{rupasov}. Our group theory formulation 
of SIT suggests an alternative, yet more systematic way of quantizing 
the SIT theory according to each specific coset structures. More 
generally, our approach can be extended to other quantum optical 
systems. Thes works are in progress and will be reported elsewhere.

\vglue .3in 
{\bf ACKNOWLEDGEMENT}
\vglue .2in
This work was supported in part by the program of Basic Science Research, 
Ministry of Education BSRI-96-2442, and by Korea Science and Engineering 
Foundation through CTP/SNU.
\vglue .2in

\end{document}